\documentclass[12pt,preprint]{aastex}
\usepackage{emulateapj5,apjfonts}

\shorttitle{X-RAY AND UV ABSORPTION IN NGC\,4051}

\shortauthors{COLLINGE ET AL.}

\journalinfo{The Astrophysical Journal, 557:1--16, 2001 August 10, astro-ph/0104125}
\slugcomment{Recieved 2001 February 5; accepted 2001 April 10}

\begin{document}

\title{High-Resolution X-ray and Ultraviolet Spectroscopy of the Complex
Intrinsic Absorption in NGC\,4051 with {\it Chandra} and {\it HST}}
\author{
M. J. Collinge,\altaffilmark{1,2} 
W. N. Brandt,\altaffilmark{1} 
Shai Kaspi,\altaffilmark{1} 
D. Michael Crenshaw,\altaffilmark{3}
Martin Elvis,\altaffilmark{4}\\
Steven B. Kraemer,\altaffilmark{3}
Christopher S. Reynolds,\altaffilmark{5,6}
Rita M. Sambruna,\altaffilmark{1,7} 
and~Beverley~J.~Wills\altaffilmark{8}
}
\altaffiltext{1}{Department of Astronomy and Astrophysics, 525 Davey
Laboratory, The Pennsylvania State University, University Park, PA, 16802 \\
({\tt collinge@astro.psu.edu, niel@astro.psu.edu, shai@astro.psu.edu, and 
rms@astro.psu.edu}).}
\altaffiltext{2}{NASA-supported undergraduate research associate.}
\altaffiltext{3}{Catholic University of America and Laboratory for 
Astronomy and Solar Physics, NASA's Goddard Space Flight Center, Code 681 
Greenbelt, MD 20771 \\ ({\tt crenshaw@buckeye.gsfc.nasa.gov and 
stiskraemer@stars.gsfc.nasa.gov}).}
\altaffiltext{4}{Harvard-Smithsonian Center for Astrophysics, 60 Garden 
Street, Cambridge, MA 02138 ({\tt elvis@head-cfa.harvard.edu}).}
\altaffiltext{5}{JILA, Campus Box 440, University of Colorado, Boulder CO 80303 
({\tt chris@rocinante.colorado.edu}).}
\altaffiltext{6}{Hubble Fellow.}
\altaffiltext{7}{Department of Physics \& Astronomy and School of Computational
Sciences, George Mason University, 4400 University Dr. M/S 3F3, 
Fairfax, VA 22030-4444.}
\altaffiltext{8}{Department of Astronomy, University of Texas at Austin, 
Austin, TX 78712 ({\tt bev@pan.as.utexas.edu}).}

%------------------------------------------------------------------------------

\begin{abstract}

We present the results from simultaneous observations of the
Narrow-Line Seyfert~1
galaxy NGC\,4051 with the {\it Chandra} High Energy Transmission
Grating Spectrometer and the {\it HST} Space Telescope Imaging
Spectrograph. The X-ray grating spectrum reveals absorption and
emission lines from hydrogen-like and helium-like ions of O, Ne, Mg and Si.
We resolve two distinct X-ray absorption systems: a high-velocity
blueshifted system at $-2340\pm 130$~km~s$^{-1}$ and a low-velocity
blueshifted system at $-600\pm 130$~km~s$^{-1}$.
In the UV spectrum we detect strong absorption,
mainly from \ion{C}{4}, \ion{N}{5} and \ion{Si}{4}, that is resolved
into as many as nine different intrinsic absorption systems with
velocities between $-650$~km~s$^{-1}$ and 30~km~s$^{-1}$.
Although the low-velocity X-ray absorption is consistent in velocity
with many of the UV absorption systems, the high-velocity X-ray
absorption seems to have no UV counterpart. In addition to the
absorption and emission lines, we also observe rapid X-ray variability
and a state of low X-ray flux during the last $\approx 15$~ks of the
observation.
NGC\,4051 has a soft X-ray excess which we fit in both the
high and low X-ray flux states. 
The high-resolution X-ray spectrum directly reveals that the soft
excess is not composed of narrow emission lines and that it has
significant spectral curvature. A power-law model fails to fit it,
while a blackbody produces a nearly acceptable fit.  We compare the
observed spectral variability with the results of previous studies of
NGC\,4051.

\end{abstract}

%-----------------------------------------------------------------------------


%--------------------------------------------------------------------------------

\section{INTRODUCTION}

NGC\,4051 is a Narrow-Line Seyfert~1 (NLS1) galaxy with a rich
observational history. Its optical emission-line spectrum was first
noted by Hubble (1926), and it is one of the archetypical Seyfert
galaxies (Seyfert 1943). It has been studied with every X-ray mission
since {\it Einstein}, as well as with {\it EUVE}, {\it HST} and {\it
IUE}, because it is bright [$V\approx13.5$; $z=0.002295\pm 0.000043$
based on optical emission lines (de Vaucouleurs et~al. 1991)] and has
relatively little Galactic absorption [$N_{\rm H}=(1.3\pm 0.1)\times
10^{20}$~cm$^{-2}$; Elvis, Lockman, \& Wilkes 1989].  NGC\,4051 is
highly variable in X-rays and has even been seen virtually to shut off 
(decreasing in flux by a factor of $\approx 20$ from the average; e.g., 
Guainazzi et~al. 1998; Uttley et~al. 1999). Its typical
2--10~keV luminosity is (2--5)$\times 10^{41}$~erg~s$^{-1}$ ($H_{\rm
0}=70$~km~s$^{-1}$~Mpc$^{-1}$; $q_{\rm 0}=0.5$). At lower energies the
X-ray continuum is dominated by a variable soft excess (e.g., Turner \&
Pounds 1989; Guainazzi et~al. 1996, hereafter G96).

NGC\,4051 is also known to contain a `warm absorber' (X-ray absorbing
highly ionized gas) from {\it ROSAT\/} and {\it ASCA\/} observations
(e.g., McHardy et~al. 1995; G96; 
Komossa \& Fink 1997; Reynolds 1997; George et~al. 1998).
Although edges from \ion{O}{7} and \ion{O}{8} have been statistically
detected in many recent analyses of NGC\,4051, the addition of these
features cannot entirely account for the observed spectral complexity
(even when a blackbody model for the soft excess is also included
in the fitting). For example, G96 identified a narrow spectral feature
near 1~keV but were unable to determine if this feature was an 
absorption edge or emission line. George et~al. (1998) suggested 
that the strong emission lines that are expected to arise in the 
warm absorber (e.g., Krolik \& Kriss 1995; Netzer 1996) may comprise 
a significant fraction of the observed soft excess and contribute 
to the spectral complexity; however, all X-ray observations to date 
have lacked the spectral resolution needed to confirm or reject this 
hypothesis. G96 also argued for significant variability of the 
\ion{O}{7} edge while the \ion{O}{8} edge remained roughly constant 
in strength, a somewhat surprising result if the \ion{O}{8} absorption 
occurs closer to the central source on average than the \ion{O}{7} 
absorption (see Reynolds 1997). The best-fit ionization parameter 
of the warm absorber in NGC\,4051 does not always track source 
intensity; such behavior could arise if the warm absorber is not 
in photoionization equilibrium or has a multi-zone nature
(e.g., McHardy et~al. 1995; Nicastro et~al. 1999). Finally, we 
note that the possible presence of intrinsic UV absorption was briefly 
suggested by Voit, Shull, \& Begelman (1987), but to date such 
absorption has not been studied in any detail.

In this paper, we present the results from our analyses of simultaneous
{\it Chandra} High Energy Transmission Grating Spectrometer (HETGS) and
{\it HST} Space Telescope Imaging Spectrograph (STIS) observations of
NGC\,4051. Through this study, we aim to constrain the dynamics (e.g.,
bulk velocity and velocity dispersion) and geometry (e.g., location and
covering factor) of the X-ray and UV absorbers, and to determine
whether these features might plausibly arise in the same gas (e.g., 
Mathur, Elvis, \& Wilkes 1995). 
As our observation comprises the first X-ray grating 
spectroscopy of NGC\,4051, we shall
also use the high spectral resolution of the HETGS to place the best
constraints to date on the spectral form and nature of the soft excess
and to examine flux and spectral variability. In \S~2 we describe the
observations and data analysis, and in \S~3 we interpret our results.

%--------------------------------------------------------------------------------

\section{OBSERVATIONS AND DATA ANALYSIS}

\subsection{{\it Chandra} HETGS Observation}
\label{cobs}

\subsubsection{Observation Details and Basic Analysis}
\label{cobs1}

NGC\,4051 was observed by {\it Chandra\/} (Weisskopf et al. 2000) on
2000 April 24--25 (starting at 06:08:25 UT) using the HETGS (C.R.
Canizares et~al. 2001, in preparation).\footnote{For additional
information on the HETGS and ACIS, see the Chandra Proposers'
Observatory Guide at http://asc.harvard.edu/udocs/docs.} 
This observation was part of the {\it Chandra\/} Cycle~1 guest observer
program (observation identification number 00859). The detector 
was the Advanced CCD Imaging
Spectrometer (ACIS; G.P. Garmire et~al. 2001, in preparation).$^9$ The
observation was continuous with a total integration time of 81.5~ks.
We used the {\it Chandra} Interactive Analysis of Observations ({\sc
ciao}) software Version 1.1.4 (M. Elvis et~al. 2001, in preparation) to
reduce the data.\footnote{See http://asc.harvard.edu/ciao.}

The zeroth order X-ray spectrum shows substantial photon pileup. 
The nuclear position derived from the {\it Chandra\/} image is 
$\alpha_{2000}=12^{\rm h}03^{\rm m}09\fs58 $,
$\delta_{2000}=+44^{\rm d}31^{\rm m}52\fs9$, 
in agreement with the radio position from Ulvestad \& Wilson (1984);
our position is offset from theirs by $0\farcs6$ to the North, while 
the absolute {\it Chandra} astrometry for our observation is 
expected to be good to within $0\farcs6$.\footnote{See 
http://asc.harvard.edu/mta/ASPECT.}
Comparing the point spread function (PSF) wings of the NGC\,4051 image
to the PSF of a point source (the HETGS observation of Capella), we
find good agreement with no clear evidence for an extended
circumnuclear X-ray component in NGC\,4051. This is in contrast to the
findings of Singh (1999) who claimed to detect extended circumnuclear
emission in a {\it ROSAT\/} HRI observation; the extended emission was
claimed to contain $21\pm 6$\% of the total flux. Our observation has
$\ga 5$ times better spatial resolution than the {\it ROSAT\/} HRI
observation and is $\approx 8$ times longer (the {\it ROSAT\/} HRI
instrument had a comparable effective area to the zeroth order of the
HETGS). The extended emission reported by Singh (1999) may be a result
of limitations in the {\it ROSAT\/} `de-wobbling' procedure or the
theoretical model of the {\it ROSAT\/} HRI PSF. We also comment on the
presence of two weak off-nuclear ($\approx 17\arcsec$ and $\approx
43\arcsec$) sources, located at 
$\alpha_{2000}=12^{\rm h}03^{\rm m}11\fs04$, 
$\delta_{2000}=+44^{\rm d}31^{\rm m}46\fs0$ and 
$\alpha_{2000}=12^{\rm h}03^{\rm m}13\fs51$, 
$\delta_{2000}=+44^{\rm d}31^{\rm m}46\fs7$, 
coincident with the galactic disk of NGC\,4051. 
These sources, which do not have any apparent discrete counterparts in
Palomar Observatory Sky Survey (POSS) or {\it HST} images, have $\approx
24$ and $\approx 11$ counts, respectively, corresponding to rest-frame
luminosities of
$L_{0.5-8 \rm\ keV}\approx 1.4\times 10^{38} \rm\ erg\ s^{-1}$ and 
$L_{0.5-8 \rm\ keV}\approx 6.6\times 10^{37} \rm\ erg\ s^{-1}$ 
if they are at the distance of NGC\,4051. They are plausibly X-ray
binaries or luminous supernova remnants in NGC\,4051, although they
could also be unrelated background sources; the probability of
obtaining one background source coincident with the optical disk of
NGC\,4051 is $\approx 40$\% at the observed flux levels.

%--------------------------------------------------------------------------------

\begin{figure*}
\centerline{\includegraphics[width=18.2cm]{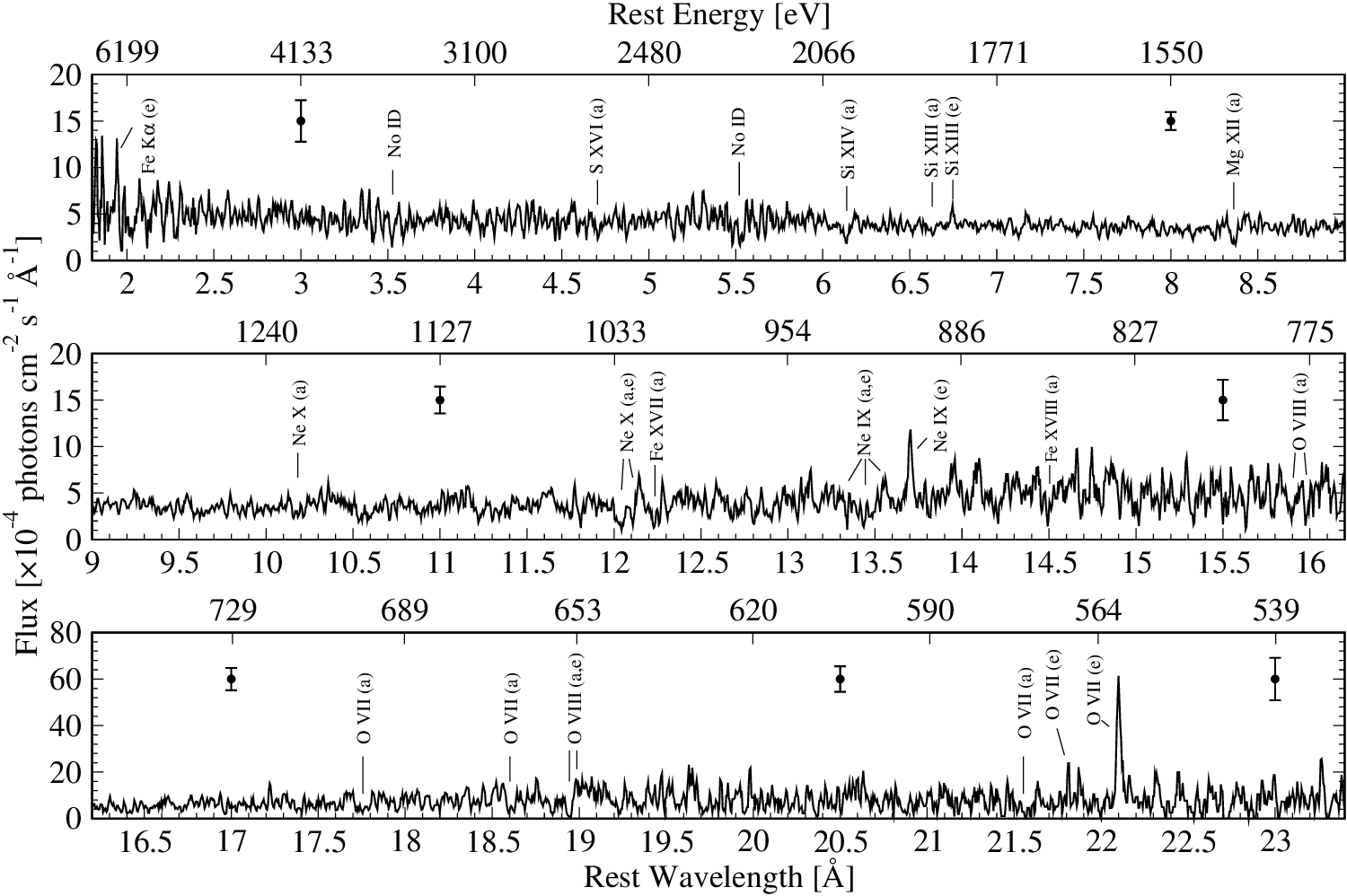}}
\caption{{\it Chandra\/} MEG first-order spectrum of NGC\,4051. The bin
size is 0.005 \AA, and the spectrum has been smoothed using a boxcar
filter that is three bins in width. Fluxes are corrected for Galactic
absorption. Dots with error bars show the typical $\pm 1 \sigma$
statistical error at various wavelengths. The identified absorption (a)
and emission (e) features are marked. Strong unidentified features are
indicated. Some additional spectral complexity appears to be present, 
particularly in the middle panel (e.g., 14--16~\AA); 
however, the wavelengths of the apparent 
features do not seem to correspond to strong expected absorption or 
emission features from either of the two identified X-ray absorption systems.
This additional complexity may be the cumulative result of a forest of 
individually weak lines (e.g., Kaspi et~al. 2001).
Note the difference in vertical scale between the top two
panels and the bottom panel.
\label{megspec}}
\end{figure*}

%--------------------------------------------------------------------------------

The HETGS produces higher order spectra from two grating assemblies, the High 
Energy Grating (HEG) and Medium Energy Grating (MEG); each 
produces positive and negative orders.
The two first-order MEG spectra agree well with each other and with the
first-order HEG spectra. The averaged first-order MEG and HEG spectra
have signal-to-noise ratios (S/Ns) of $\approx 4.6$ and $\approx 2.3$
(near 7~\AA) for bin sizes of 0.005~\AA\ and 0.0025~\AA,
respectively. The higher orders have only a small number of counts, and
we therefore exclude them from our analysis. We have used {\sc ciao} to
produce Ancillary Response Files (ARFs) in order to flux calibrate the
spectra, and we have corrected for the Galactic absorption and the
cosmological redshift. The flux calibration is currently estimated to
be accurate to $\approx 30$\%, 20\% and 10\% in the 0.5--0.8, 0.8--1.5
and 1.5--6~keV bands, respectively (H. L. Marshall 2000, private
communication).\footnote{See
http://space.mit.edu/ASC/calib/hetgcal.html.} We have also corrected
for the wavelength shift of 0.054\% induced by the thermal contraction
of ACIS-S pixels (H. L. Marshall 2000, private communication). We
present the mean MEG first-order spectrum in Figure~\ref{megspec}. This
spectrum contains a total of 32087~counts between 1.8~\AA\ and 25~\AA\
(0.5--6.9 keV). The MEG velocity resolution is 1400~km\,s$^{-1}$ at
5~\AA\ and 350~km\,s$^{-1}$ at 20~\AA .

The general shape and flux of the X-ray spectrum are consistent with
previous observations of NGC\,4051.  The average fluxes measured from
the {\it Chandra} spectrum are ($1.4\pm0.3$)$\times
10^{-11}$~erg~cm$^{-2}$~s$^{-1}$ in the 0.5--2~keV band and
($1.2\pm0.1$)$\times 10^{-11}$~erg~cm$^{-2}$~s$^{-1}$ in the 2--7~keV
band; the errors on these figures represent the flux calibration
uncertainties mentioned above. We also observe short-timescale
variability (see Figure~\ref{curve}) and a state of low flux during the
last 15~ks of the observation. For this interval, the average {\it
Chandra} fluxes are ($2.9\pm0.6$)$\times
10^{-12}$~erg~cm$^{-2}$~s$^{-1}$ in the 0.5--2~keV band and
($5.5\pm0.6$)$\times 10^{-12}$~erg~cm$^{-2}$~s$^{-1}$ in the 2--7~keV
band; these fluxes, while quite low for NGC\,4051, are still about four
times higher than during the most extreme of the ultra-dim X-ray states
mentioned in \S~1.
The largest amplitude of count-rate variability observed is a factor of
$12.6\pm 0.7$, and the average count rate during the last 15~ks of the
observation is $4.8\pm 0.1$ times smaller than for the preceding data.
We note that the low-flux state could not have lasted for more
than $\approx 3$~days, as NGC\,4051 was observed to be in a more
typical flux state shortly after our observation (P. Uttley 2000,
private communication).

\subsubsection{Detected X-ray Spectral Features}
\label{cobs2}

The most interesting features revealed in the high-resolution X-ray
spectrum of NGC\,4051 are narrow absorption and emission lines, similar
to those recently detected in other Seyfert~1s such as NGC\,3783 (Kaspi
et~al. 2000) and NGC\,5548 (Kaastra et~al. 2000). Because only a small
number of the individual 
%--------------------------------------------------------------------------------
\centerline{\includegraphics[width=8.5cm]{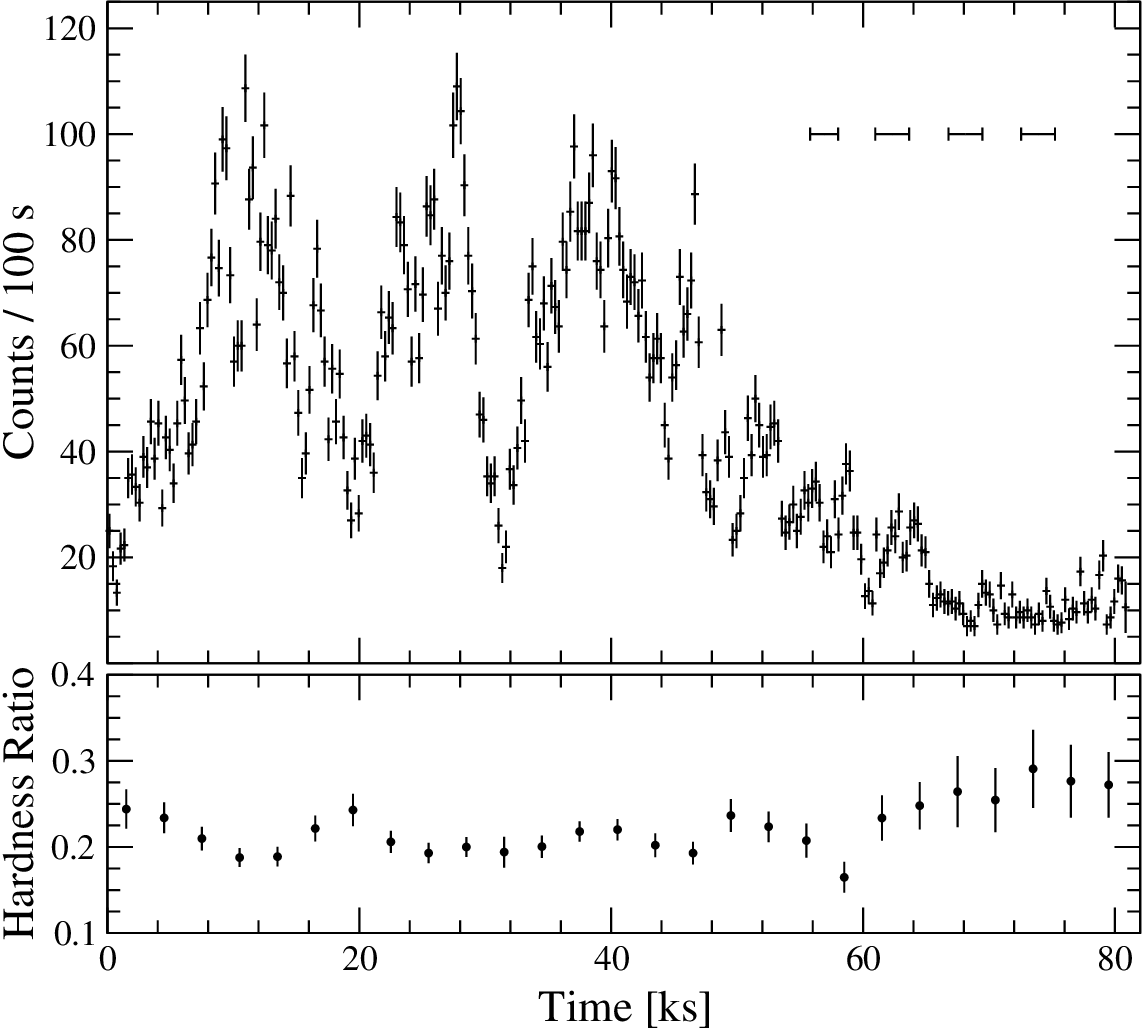}}
\figcaption{{\it Chandra\/} MEG light curve and hardness ratio for
NGC\,4051. The hardness ratio is defined to be the ratio of the
2--8~keV and 0.5--2~keV counts.  The top panel has a bin size of 300~s,
and the bottom panel has a bin size of 3000~s. Note the low count rate
during the last $\approx 15$~ks; the spectrum clearly hardens during
this time. The horizontal lines in the top panel show the times when
STIS data were acquired.
\label{curve}}
\centerline{}
%--------------------------------------------------------------------------------

\noindent
absorption features are statistically
significant at a high level of confidence, we have adopted the
following procedure to measure relevant properties of the lines.
We created `velocity spectra' by adding, in velocity space,
several absorption lines from the same ion. 
This method has the advantage of a significant increase in S/N and the
disadvantage of somewhat deteriorated spectral resolution since, for
each ion, the resolution is then defined by the line with the
shortest wavelength.
The velocity spectra were 
built up on a photon-by-photon basis from the
raw data, rather than by interpolating spectra already binned in
wavelength. 
In all cases, the two strongest predicted features for each ion were
included (H. Netzer 2000, private communication), and in some cases
other lines were added in order to improve the strength of the signal.
We chose the lines to be relatively free from contamination by
unrelated adjacent features based upon observations and modeling of
other Seyfert galaxies with warm absorbers (mainly NGC\,3783; Kaspi
et~al. 2001).
In addition to combining the absorption lines from single
ions, we also combined all the transitions from H-like ions into one
spectrum and all the transitions from He-like ions into another. This
step was phenomenologically motivated by the observed similarities
between many of the spectra for the H-like and He-like ions. The
velocity spectra created in \  this \ way \ are \ shown

%--------------------------------------------------------------------------------
\begin{figure*}
\centerline{\includegraphics[width=18.5cm]{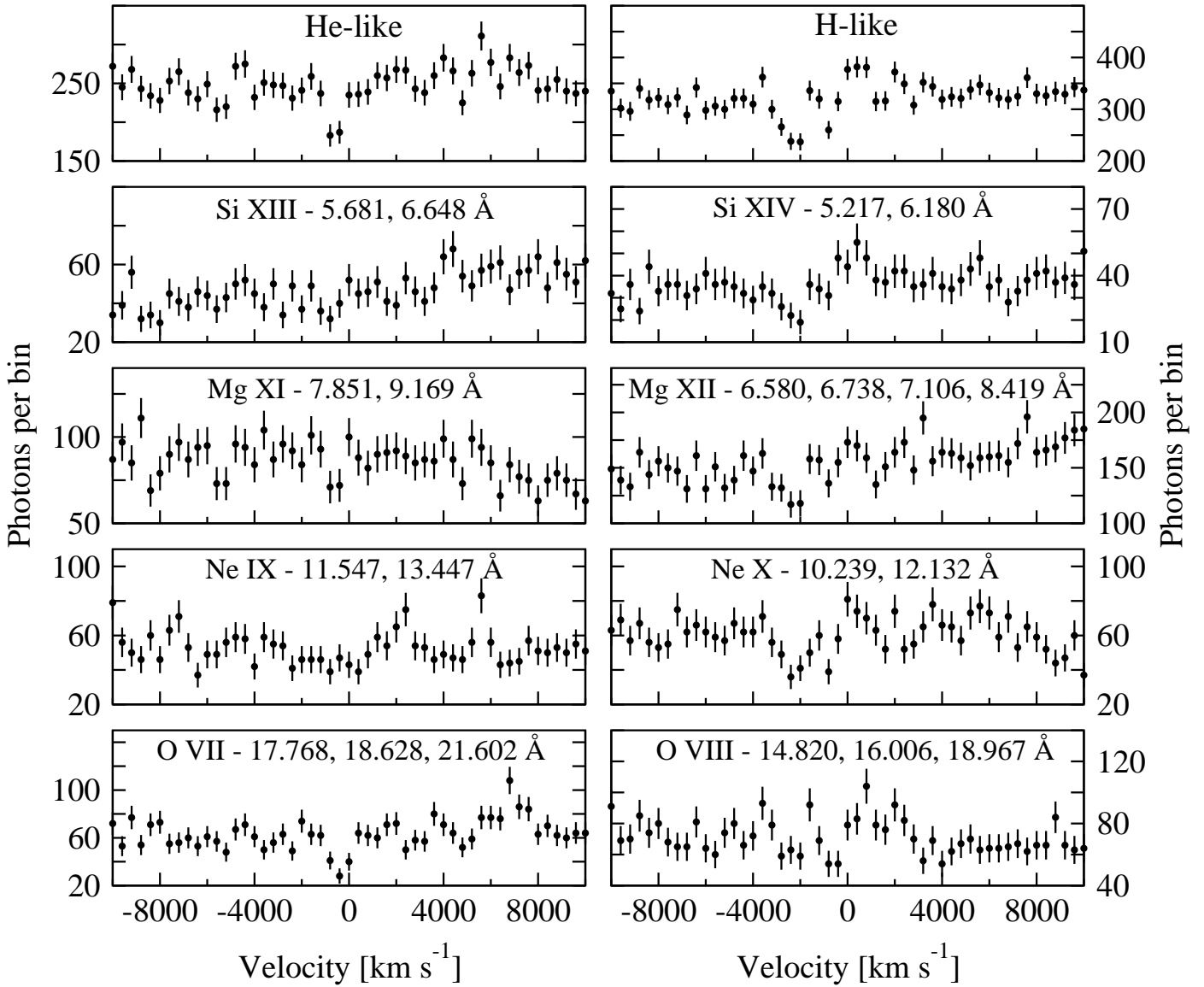}}
\figcaption{{\it Chandra\/} MEG velocity spectra showing coadded lines
from H-like and He-like ions of O, Ne, Mg and Si. The bin size is
400~km~s$^{-1}$. The wavelengths of the lines coadded in each spectrum
are listed in the respective panels, except for the total H-like and
He-like spectra which contain all of the lines listed for that type of ion. 
Absorption systems at 
two distinct velocities ($-2340$~km~s$^{-1}$ and $-600$~km~s$^{-1}$)
are detected when all ions are considered (compare the upper two panels).
\label{profiles}}
\end{figure*}
%--------------------------------------------------------------------------------

\noindent
in Figure~\ref{profiles}. We measured the velocity shifts and widths of
the observed absorption systems from Gaussian fits to the total H-like
and He-like spectra. We detect two distinct absorption components with
$>99.9$\% confidence according to the $F$-test (see Bevington \&
Robinson 1992): a high-velocity system at $-2340\pm 130$~km~s$^{-1}$
($\Delta \chi^2=59.70$ in the H-like spectrum) and a low-velocity
system at $-600\pm 130$~km~s$^{-1}$ ($\Delta \chi^2=38.34$ in the
He-like spectrum and $\Delta \chi^2=15.41$ in the H-like spectrum). 
The two absorption components have FWHMs of $980\pm
310$~km~s$^{-1}$ and $770\pm 460$~km~s$^{-1}$, respectively. These
values are consistent with the resolution of the MEG at the wavelengths
of some of the lines in each coadded spectrum.

Constraining the velocity shifts of the two components to have the
values we measured from the H-like (high-velocity component) and
He-like (low-velocity component) ions, we then performed $F$-tests on
the velocity spectra of the individual ions in order to check
statistically for the presence of both absorption systems from each
ion. This process consisted of fitting the continuum from $-5000$ to
5000~km~s$^{-1}$ with a linear model and noting the change in $\chi^2$
when narrow (unresolved) Gaussian absorption lines were added at fixed
velocities corresponding to either high-velocity or low-velocity
absorption. Because the only free parameter in the Gaussian model was
the normalization, the values of $\Delta \chi^2$ correspond to one
parameter of interest. The results of this procedure are shown in
Table~1. According to the $F$-test, we detect absorption at
the $>99.9$\% confidence level from the high-velocity system in
\ion{Si}{14}, \ion{Mg}{12}, and \ion{Ne}{10}, and from the low-velocity
system in \ion{O}{7}.  Absorption is also suggested 
from the remaining H-like or He-like species of O, Ne,
Mg and Si. Because deviations in the continuum from a linear model may introduce 
small inaccuracies into our $F$-test results, we only consider 
features indicated at 
the $>99.9$\% confidence level to be reliable detections.

\newpage

%--------------------------------------------------------------------------------

%\end{multicols}
%\setcounter{table}{0}
%\begin{deluxetable}{lcc}
%\tabletypesize{\footnotesize}
%\tablecolumns{3}
%\tablewidth{0pt}
%\tablecaption{\sc Values of $\Delta \chi^2$ for Co-added X-ray Line Profiles
%\label{coadd}}
%\tablehead{
%
\footnotesize % better fit the Journal font size.
\begin{center}
{\sc TABLE 1 \\  Values of $\Delta \chi^2$ for Co-added X-ray Line Profiles}
\vskip 4pt
\begin{tabular}{lcc}
\hline
\hline
{} &
{ \ \ \  \ \ \  \ \ \ Low Vel. \ \ \  \ \ \  \ \ \ } &
{ \ \ \  \ \ \  \ \ \ High Vel. \ \ \  \ \ \  \ \ \ } \\
{ \ \ \   \ \ \ Ion  \ \ \  \ \ \  \ \ \ } &
{$\Delta \chi^2$} &
{$\Delta \chi^2$} \\ [0.1cm]
\hline
\ion{Si}{13} & \phn 4.66 & \phn 0.38 \\ 
\ion{Si}{14} & \phn 0.49 & 14.75     \\ 
\ion{Mg}{11} & \phn 4.72 & \nodata  \\ 
\ion{Mg}{12} & \phn 0.38 & 18.05     \\ 
\ion{Ne}{9}  & \phn 5.82 & \phn 5.77 \\ 
\ion{Ne}{10} & \phn 3.44 & 21.19     \\ 
\ion{O}{7}   &     37.72 & \phn 0.49\tablenotemark{*} \\ 
\ion{O}{8}   & \phn 6.79 & \phn 3.44\tablenotemark{*} \\ 
\hline
\end{tabular}
\vskip 2pt
\parbox{3.2in}{ 
\small\baselineskip 9pt
\footnotesize
\indent
{ NOTE. ---
Values of $\Delta \chi^2$ represent the difference in $\chi^2$ derived from 
linear models and linear models with one Gaussian component to represent 
absorption from the indicated velocity system, except for the values marked with 
`*'. In these cases, $\Delta \chi^2$ is the difference in $\chi^2$ derived from 
linear models with one Gaussian component to represent low-velocity absorption 
and linear models with Gaussian components to represent absorption at both 
velocities. All fits have at least 40 degrees of freedom. For 40 degrees of 
freedom, some $\Delta \chi^2$ values and their corresponding significance levels 
are: $\Delta \chi^2=4.08$ / 95\%; $\Delta \chi^2=7.31$ / 99\%; 
$\Delta \chi^2=12.6$ / 99.9\% (see Table~C.5 of Bevington \& Robinson 1992).}
}
\end{center}
\setcounter{table}{1}
\normalsize
\centerline{} % just needed to provide some space.
\centerline{} % just needed to provide some space.

%--------------------------------------------------------------------------------

In Table~2 we list information about the individual
absorption and emission lines. We have used the same procedure detailed
in the preceding paragraph to search for individual absorption lines in
the X-ray spectrum, again constraining the absorption components to be
narrow and to occur at wavelengths corresponding to the velocities
given above. Although this procedure was very effective at identifying 
strong features, we note that it cannot entirely account for 
the observed spectral complexity (see Figure~\ref{megspec} for 
clarification).
In regions of the spectrum where there are not enough
counts to permit $\chi^2$ fitting, we have manually searched for
absorption features. Similarly, we have not performed $\chi^2$
statistical significance testing on any of the identified emission
lines because many of these occur in regions of the spectrum with low
S/N.  We measured the velocity shifts and widths of the emission lines
from simple Gaussian fits. Where allowed by the strength of the signal,
we have calculated the errors on these values at the 90\% confidence
level ($\Delta \chi^2=2.71$). In other cases, we have estimated the
errors by visual inspection of the features. We note that the emission
lines appear to be unresolved (see \S~3.1 for further discussion) and
are generally consistent with zero velocity within the
NGC\,4051 rest frame. We present the X-ray emission-line fluxes in 
Table~3.
In general, the statistical significances of the emission lines and the
low S/N absorption lines can be judged by the sizes of the errors on
the equivalent widths (EWs). In order to measure the EWs of all the
individual features, we have excised the line-dominated areas in the
spectrum and fit the remainder with a cubic spline which serves as the
continuum reference point in the centers of the lines. Errors on EWs
represent the combination of a global 10\% uncertainty in the continuum
level and the statistical photon noise.

Some of the strong absorption lines occurring at the
long-wavelength end of the MEG spectrum appear to be resolved (see
Figure~\ref{resolved}). Fitting a Gaussian profile to the coadded
\ion{O}{7} velocity spectrum yields a best-fit FWHM for the
low-velocity absorption  system of $850^{+380}_{-240}$~km~s$^{-1}$
(errors are for $\Delta \chi^2=2.71$), whereas the instrumental
resolution (FWHM) of the MEG at 17.768~\AA\ (the line in the \ion{O}{7}
velocity spectrum with the shortest wavelength and therefore the worst
resolution) is 390~km~s$^{-1}$. Correcting for the instrumental
broadening gives a true FWHM of $750^{+400}_{-280}$~km~s$^{-1}$ for the
low-velocity X-ray absorption system. Although the high-velocity system
appears to

%--------------------------------------------------------------------------------

%\end{multicols}
%\begin{deluxetable}{lcccc}
%\tabletypesize{\footnotesize}
%\tablecolumns{5}
%\tablewidth{0pt}
%\tablecaption{\sc X-ray Absorption and Emission Lines From NGC\,4051
%\label{xlines}}
%\tablehead

\footnotesize % better fit the Journal font size.
\begin{center}
{\sc TABLE 2 \\  X-ray Absorption and Emission Lines From NGC\,4051}
\vskip 4pt
\begin{tabular}{lcccc}
\hline
\hline
{} &
{Rest} &
{} &
{Velocity} &
{Measured} \\
{} &
{$\lambda$} &
{} &
{(System} &
{EW} \\
{Ion} &
{(\AA)} &
{$\Delta \chi^2$} &
{or km s$^{-1}$)} &
{(m\AA)} \\
\hline
\multicolumn{5}{c}{Absorption Lines} \\
\hline
\ion{S }{16}                &  \phn4.728 & \phn2.20 & High &  $7.0^{+5.2}_{-5.1\phn}$ \\ 
\ion{Si}{14}                  &  \phn6.181 & 13.05 & High &  $13.0^{+4.4}_{-4.0\phn\phn}$ \\ 
\ion{Si}{13}                  &  \phn6.648 & \phn3.58 & Low &  $5.3^{+2.7}_{-2.6\phn}$ \\ 
\ion{Mg}{12}\tablenotemark{a} &  \phn8.419 & 14.14 & High &  $18.7^{+3.9}_{-3.6\phn\phn}$ \\ 
\ion{Ne}{10}\tablenotemark{a} & 10.239 & \phn2.90 & High &  $6.2^{+2.4}_{-2.3\phn}$ \\ 
\ion{Ne}{10}\tablenotemark{a} & 12.132 & \phn4.60 & Low &  $10.9^{+4.1}_{-4.0\phn\phn}$  \\ 
\ion{Ne}{10}\tablenotemark{a} & 12.132 & 11.52 & High &  $35.9^{+6.5}_{-6.1\phn\phn}$  \\ 
\ion{Fe}{17}\tablenotemark{a} & 12.263 & \phn7.30 & Low & $16.2\pm 4.0\phn$ \\ 
\ion{Ne}{ 9}                  & 13.447 & \phn4.70 & Low & $15.9^{+5.1}_{-5.0\phn\phn}$   \\
\ion{Ne}{ 9}                  & 13.447 & \phn2.90 & High &  $11.0^{+3.2}_{-3.1\phn\phn}$  \\
\ion{Fe}{18}                  & 14.539 & \phn2.33 & Low & $9.9^{+4.1}_{-4.0\phn}$  \\
\ion{O }{ 8}\tablenotemark{a} & 16.006 & \phn3.48 & Low &  $10.3\pm 3.2\phn$  \\ 
\ion{O }{ 8}\tablenotemark{a} & 16.006 & \phn2.93 & High &  $\phn6.3\pm 3.6\phn$  \\ 
\ion{O }{ 7}                  & 17.768 & \phn3.15 & Low &  $17.5\pm 8.5\phn $ \\ 
\ion{O }{ 7}                  & 18.628 & \phn6.28 & Low &  $24.8\pm 9.5\phn $ \\
\ion{O }{ 8}                  & 18.967 & \nodata & Low &  $30.1\pm 4.9\phn$  \\ 
\ion{O }{ 7}                  & 21.602 & \nodata & Low &  $37.1\pm 12.2$ \\ 
\hline
\multicolumn{5}{c}{Emission Lines} \\
\hline
\ion{Fe}{1}\,--\ion{}{18}     &  \phn1.937 & & $-250\pm 750$\tablenotemark{b}  &  $\phn47.7^{+15.4}_{-14.2}$ \\
\ion{Si}{13}                  &  \phn6.741 & & $\phn\phn\phn0^{+250}_{-310}$  &  $\phn12.0^{+6.4\phn}_{-5.6\phn}$ \\
\ion{Ne}{10}                  & 12.132 & & $\phn210^{+220}_{-150}$  &  $\phn15.1^{+6.7\phn}_{-6.4\phn}$ \\ 
\ion{Ne}{ 9}                  & 13.553 & & $-50^{+610}_{-590}$  &  $\phn\phn9.7^{+6.7\phn}_{-6.5\phn}$ \\ 
\ion{Ne}{ 9}                  & 13.699 & & $-63^{+225}_{-212}$  &  $\phn48.6^{+14.6}_{-13.5}$ \\ 
\ion{O }{ 8}                  & 18.967 & & $\phn180^{+120}_{-120}$  &  $\phn21.4^{+9.9\phn}_{-9.6\phn}$ \\ 
\ion{O }{ 7}                  & 21.807 & & $-60^{+100}_{-100}$  &  $\phn37.3^{+21.6}_{-21.3}$ \\ 
\ion{O }{ 7}                  & 22.102 & & $-122^{+125}_{-114}$  &  $180.1^{+45.2}_{-42.9}$ \\ 
\hline
\end{tabular}
\vskip 2pt
\parbox{3.in}{    % use this to define the width of the notes under the table
\small\baselineskip 9pt
\footnotesize
\indent
$\rm ^a${Possibly blended with Fe absorption lines.} \\
$\rm ^b${The velocity of Fe~K$\alpha$ is not well-constrained
due to the low S/N in that region of the spectrum (see Figure
\ref{emission}). The velocity reported here is the average velocity
measured in the MEG and HEG spectra.
The wavelength of this observed line corresponds to the expected iron
K$\alpha$ line from \ion{Fe}{1} to \ion{Fe}{18}.}
}
\end{center}
\setcounter{table}{2}
\normalsize

%--------------------------------------------------------------------------------
%--------------------------------------------------------------------------------

%\end{multicols}
%\begin{deluxetable}{lcc}
%\tabletypesize{\footnotesize}
%\tablecolumns{3}
%\tablewidth{0pt}
%\tablecaption{\sc X-ray and UV Emission-Line Fluxes 
%\label{emission_flux}}
%\tablehead{
\footnotesize % better fit the Journal font size.
\begin{center}
{\sc TABLE 3 \\ X-ray and UV Emission-Line Fluxes}
\vskip 4pt
\begin{tabular}{lcc}
\hline
\hline
{} &
{Rest} &
{} \\
{} &
{Wavelength} &
{Flux} \\
{Ion} &
{(\AA)} &
{($\times 10^{-14}$ erg cm$^{-2}$ s$^{-1}$)} \\
\hline
\multicolumn{3}{c}{X-ray Emission Lines} \\
\hline
\ion{Fe}{1}\,--\ion{}{18} & \phn1.937 & $32^{+11}_{-10}$  \\
\ion{Si}{13}              & \phn6.741 & $1.3^{+0.7}_{-0.6}$  \\
\ion{Ne}{10}              & 12.132 & $1.0\pm 0.4$  \\
\ion{Ne}{9}               & 13.553 & $0.6\pm 0.4$  \\
\ion{Ne}{9}               & 13.699 & $2.9^{+0.9}_{-0.8}$  \\
\ion{O}{8}                & 18.967 & $1.6\pm 0.7$  \\
\ion{O}{7}                & 21.807 & $2.4\pm 1.4$  \\
\ion{O}{7}                & 22.102 & $11\pm 3$  \\
\hline
\multicolumn{3}{c}{UV Emission Lines} \\
\hline
\ion{H}{1}                & 1215.7 & 137\tablenotemark{a}  \\
\ion{N}{5}                & 1238.8/1242.8 & 24\tablenotemark{a}  \\
\ion{Si}{2}               & 1304.4 & $12\pm 3$  \\
\ion{C}{2}                & 1334.5 & $3.1\pm 0.5$  \\
\ion{Si}{4}               & 1393.8/1402.8 & $21\pm 5$  \\
\ion{N}{3}                & 1485.8 & $3.0\pm 0.7$  \\
\ion{C}{4}                & 1548.2/1550.8 & 117\tablenotemark{a}  \\
\ion{He}{2}               & 1639.8 & $18\pm 5$  \\
\hline
\end{tabular}
\vskip 2pt
\parbox{2.8in}{    % use this to define the width of the notes under the table
\small\baselineskip 9pt
\footnotesize
\indent
$\rm ^a${Fluxes for the heavily absorbed UV lines, Ly$\alpha$, \ion{N}{5}, 
and \ion{C}{4}, are highly uncertain as a result of the strong absorption.}
}
\end{center}
\setcounter{table}{3}
\normalsize
\centerline{} % just needed to provide some space.

%--------------------------------------------------------------------------------

\noindent
be unresolved in all of the coadded line profiles, we can
use the \ion{Ne}{10} profile to place an upper limit on its measured
FWHM of 1140--1860~km~s$^{-1}$, which corresponds to a true width of
$\la$910--1730~km~s$^{-1}$. We are unable to use the
HEG spectrum to place tighter constraints on the lines' widths due to
its low S/N.

%--------------------------------------------------------------------------------

\end{multicols}
\begin{deluxetable}{lcccc}
\tabletypesize{\footnotesize}
\tablecolumns{5}
\tablewidth{0pt}
\tablecaption{\sc $\chi^2$/{\it d.o.f.} (Degrees of Freedom) for X-ray Spectral 
Fitting
\label{chivals}}
\tablehead{
\colhead{} &
\colhead{Power law} &
\colhead{Power law and} &
\colhead{Power law and} \\
\colhead{} &
\colhead{($>2$~keV data only)} &
\colhead{power law} &
\colhead{blackbody}}
\startdata
Low State & 37.1 / 42 & 137.7 / 103 & 133.9 / 103 \\
High State & 523.7 / 585 & 2178.0 / 1942 & 2060.3 / 1942 \\
Average Spectrum & 593.3 / 623 & 2255.0 / 2019 & 2133.7 / 2019 \\
\enddata
\end{deluxetable}
\begin{multicols}{2}

%--------------------------------------------------------------------------------

%--------------------------------------------------------------------------------

%\begin{figure*}
\centerline{\includegraphics[width=8.5cm]{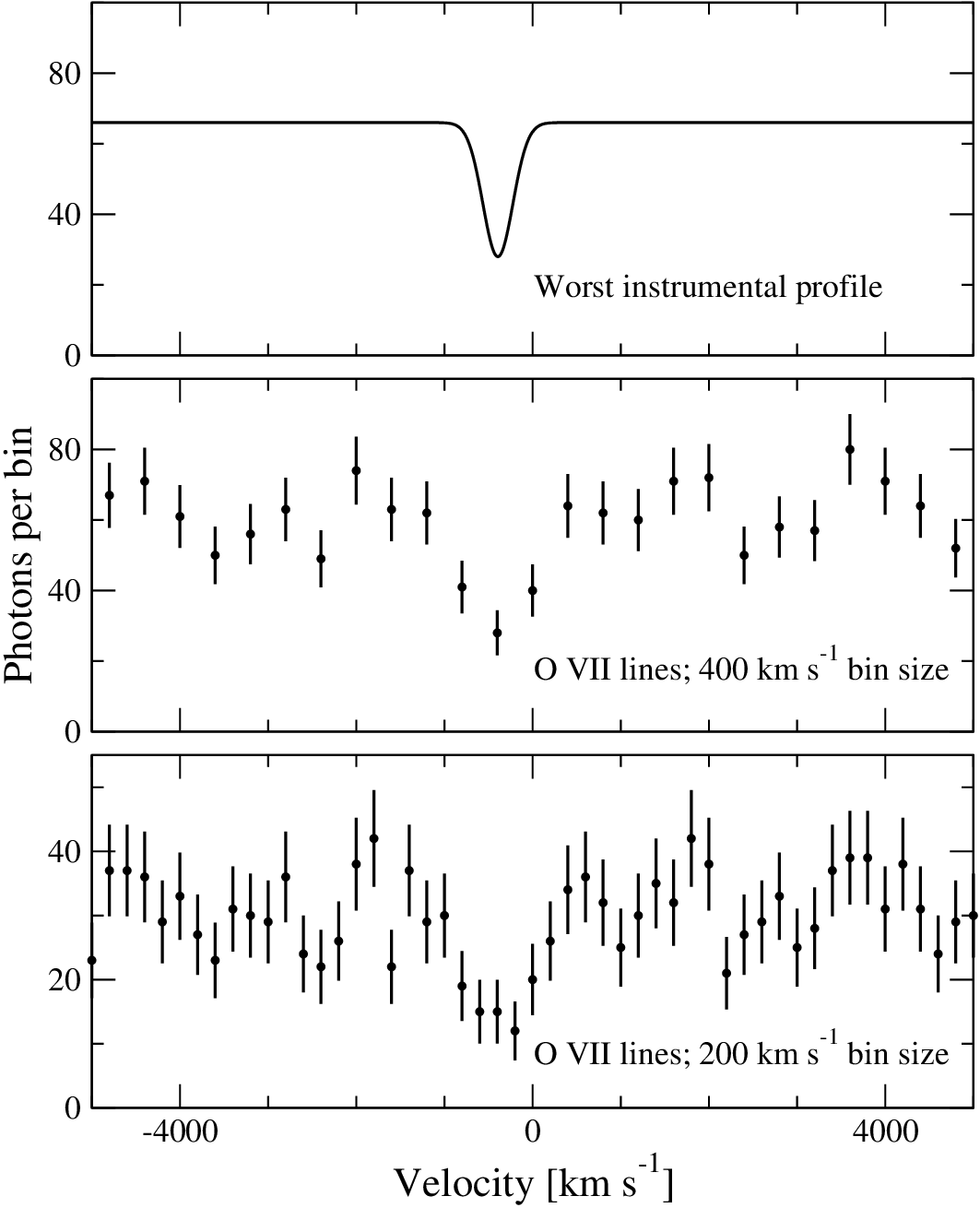}}
\figcaption{\ion{O}{7} velocity spectrum from Figure~\protect\ref{profiles} built
up from the 17.768~\AA, 18.628~\AA\ and 21.602~\AA\ lines. The top
panel shows a simulated unresolved absorption line profile as it would
appear in the MEG at a wavelength of 17.768~\AA\ (the shortest
wavelength \ion{O}{7} line and therefore the one with the worst MEG
velocity resolution).  The coadded spectrum is shown with two different
bin sizes to allow easy comparison with the instrumental response.
\label{resolved}}
%\end{figure*}
\centerline{}
\centerline{}

%--------------------------------------------------------------------------------

We have checked the two strongest emission lines, \ion{O}{7}
22.102~\AA\ and \ion{Ne}{9} 13.699~\AA , for flux variability between
the high and low states (see \S~\ref{cobs3}) and find no clear evidence
for it. The upper limit on the variability amplitude of the \ion{O}{7}
line, which is stronger than the \ion{Ne}{9} line, is $\approx 51$\%.

\subsubsection{Spectral Variability and Soft X-ray Excess}
\label{cobs3}

We have independently analyzed the grating spectra from the first
$\approx 65$~ks of the observation (hereafter the `high state') and the
last $\approx 15$~ks (`low state') in order to study spectral
variability (see Figure~\ref{curve}). We have binned the spectra to
have a minimum of 15 counts per data point, in order to allow $\chi^2$
fitting.  Since we have flux calibrated the spectra and corrected for
Galactic absorption and redshift, we have used the {\sc ciao} tool {\sc
sherpa} to fit spectral models directly to the data, without having to
make further accounting for instrumental response.  We simultaneously
fit all four first-order HETGS spectra, constraining all shared
spectral shape parameters to be the same, but allowing the absolute
model normalizations to vary (to allow for absolute instrumental 
flux-calibration uncertainties). We note that the model normalizations were
in general consistent to within 10\%. In order to obtain the most
reliable constraints on continuum spectral properties, we have excised
all strong narrow features prior to fitting. In the high-state and average
spectra, we have removed all the lines identified in Table~2
as well as the unidentified features marked in Figure~\ref{megspec}. In
the low state, we have removed only those parts of the spectrum
containing strong emission lines.

\begin{figure*}
\centerline{\includegraphics[width=12cm]{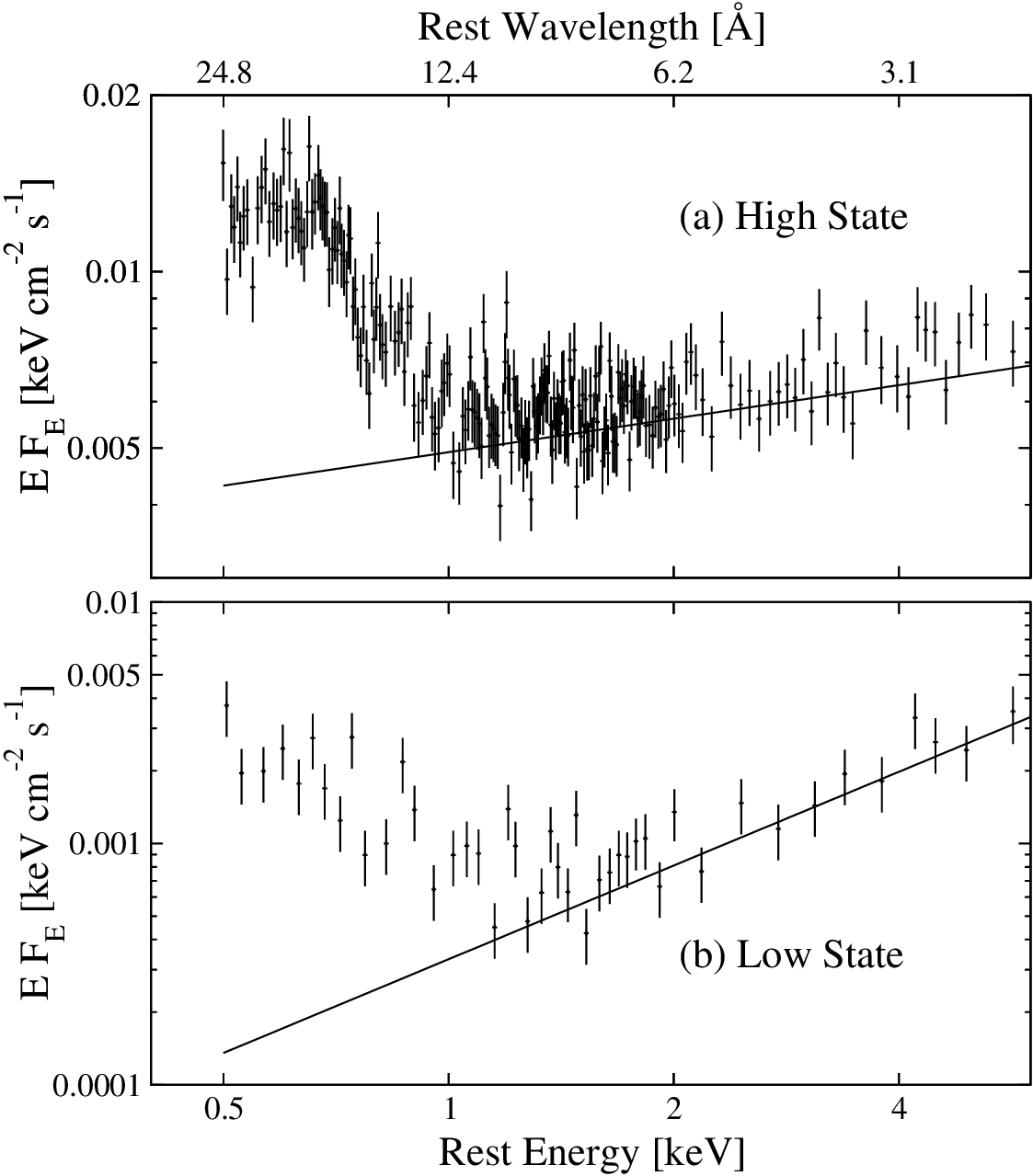}}
\caption{(a) High-state and (b) low-state MEG negative first-order
spectra of NGC\,4051. Each data point represents a minimum of 60 counts
in the top panel and a minimum of 15 counts in the bottom panel. Strong
absorption and emission lines have been excised from the spectra. We
have fit the continua above 2~keV with power laws (using all four
first-order HETGS spectra; see \S2.1.3), which we have extrapolated to
lower energies to show the soft excess. 
Note the significant spectral curvature of the soft excess apparent in the
top panel (at $\la 0.7$~keV).
\label{soft_excess}}
\end{figure*}

We fit the $>2$~keV continuum with a power law. The fits were
statistically acceptable (see Table~\ref{chivals}), and the best-fit
photon indices with 90\% confidence ranges ($\Delta \chi^2=2.71$) were
$1.85\pm0.05$ for the high state and $0.83^{+0.18}_{-0.19}$ for the low
state. The flattening of the hard power-law continuum in the low state
is qualitatively consistent with the behavior seen in previous X-ray
observations (see \S3.2 for discussion). Upon extrapolating the
power-law fits to lower energies, the soft excess becomes strongly
apparent (see Figure~\ref{soft_excess}). It appears to be continuous
rather than composed of many narrow emission lines. 
Given the detection of this strong and
continuous soft excess in high-resolution X-ray data, it seems unlikely
that the extreme-ultraviolet to hard X-ray continuum of NGC\,4051 is
composed only of a single power law (see Uttley et~al. 2000).
Comparison of the two panels of Figure~\ref{soft_excess} shows that the 
soft excess is rapidly variable. In order to measure the amplitude of this 
variability, we have subtracted the power-law models shown in 
Figure~\ref{soft_excess} from the data and calculated the remaining 
0.5--1.5~keV fluxes in both the high and low states. The ratio of the high-state 
to low-state soft excess fluxes we obtain is $2.56^{+1.06}_{-0.82}$. 
The soft excess is proportionally stronger (relative to the underlying 
continuum) in the low state than in the high state; the fraction of the 
0.5--1.5~keV flux due to the soft excess is $\approx 76$\% in the low state 
and $\approx 36$\% in the high state. 
We comment that our basic results on
the soft excess are not sensitive to the correction for
Galactic absorption; NGC\,4051 has a small and precisely measured
Galactic column density (see \S1).

We first attempted to fit the soft excess with a power law. This model
is rejected at $>99.9$\% confidence for the high state and at 98.7\%
confidence for the low state (see Table~\ref{chivals}). In the high
state, this fit displays systematic residuals in the soft band, indicating 
that the soft excess has significant spectral curvature. 
Therefore we next attempted to fit it
with a curved model, a blackbody. This provides a highly significant
improvement to the fit in the high state ($\Delta \chi^2=117.7$), while
in the low state the fit improves only slightly ($\Delta \chi^2=3.8$);
the shape of the soft excess is poorly constrained in the low state due
to its low S/N. The addition of a second blackbody component with a
different temperature does not result in a statistically significant
improvement in either state. Although the single blackbody model is
still statistically rejected with high confidence, this is largely the result of 
unmodeled spectral \ complexity, \ much \ of \ which

%--------------------------------------------------------------------------------

%--------------------------------------------------------------------------------

\noindent
may be from iron
L-shell lines that are individually below the detection threshold but
which have a significant cumulative effect (e.g., Kaspi et~al. 2001).
The spectral shape of the blackbody component is consistent to first
order with the shape of the soft excess. The best-fit blackbody
temperatures (with 90\% confidence uncertainties) for the high and low 
states are $0.1045\pm0.0006$~keV and
$0.1126\pm0.0019$~keV, respectively. Adding an intrinsic neutral
absorption column does not improve the fit; the upper limit on the
column density of such a component is $\approx 10^{20}$~cm$^{-2}$.  We
note that the presence of any intrinsic, neutral absorption in NGC\,4051 would
only increase the strength of the required soft X-ray excess.

Adopting the blackbody as the best model for the soft-excess continuum, 
we then statistically tested for the presence of absorption edges in 
the high-state spectrum. We find evidence for an 
\ion{O}{7} absorption edge ($\Delta \chi^2=25.5$) with a velocity 
consistent with the range of velocities encompassed by the X-ray 
absorption lines and with $\tau = 0.220\pm0.035$, implying an \ion{O}{7} 
column density 
of $(8\pm 1)\times 10^{17}$~cm$^{-2}$. The precise energy of 
the edge ($0.741\pm0.019$~keV) is not well 
constrained due to the loss of resolution necessary to reach $\chi^2$ 
statistics. Our test results also suggest the presence of an 
\ion{O}{8} edge ($\Delta 
\chi^2=17.1$) with a similar and equally uncertain implied velocity 
(energy in the range 0.8714--0.8787~keV) and with 
$\tau = 0.157\pm0.032$, implying an \ion{O}{8} column density of 
$(1.4\pm 0.3)\times 10^{18}$~cm$^{-2}$. In 
addition, there is statistical evidence for another absorption edge
with an energy of $\approx 0.91$~keV and $\tau\approx 0.17$ ($\Delta \chi^2 
\approx 20$; the 
energy, strength and significance depend on the presence or absence of other
edges in the model). This feature may be caused by the previously
mentioned iron L-shell absorption lines. We note that the presence of such 
absorption lines may also result in systematic 
uncertainties in the \ion{O}{7} and \ion{O}{8} edge depths and hence the implied 
column densities.

%--------------------------------------------------------------------------------

\subsection{{\it HST} STIS Observation}
\label{stobs}

\subsubsection{Observation Details and Basic Analysis}
\label{stobs1}

%--------------------------------------------------------------------------------

\begin{figure*}
\centerline{\includegraphics[width=18.5cm]{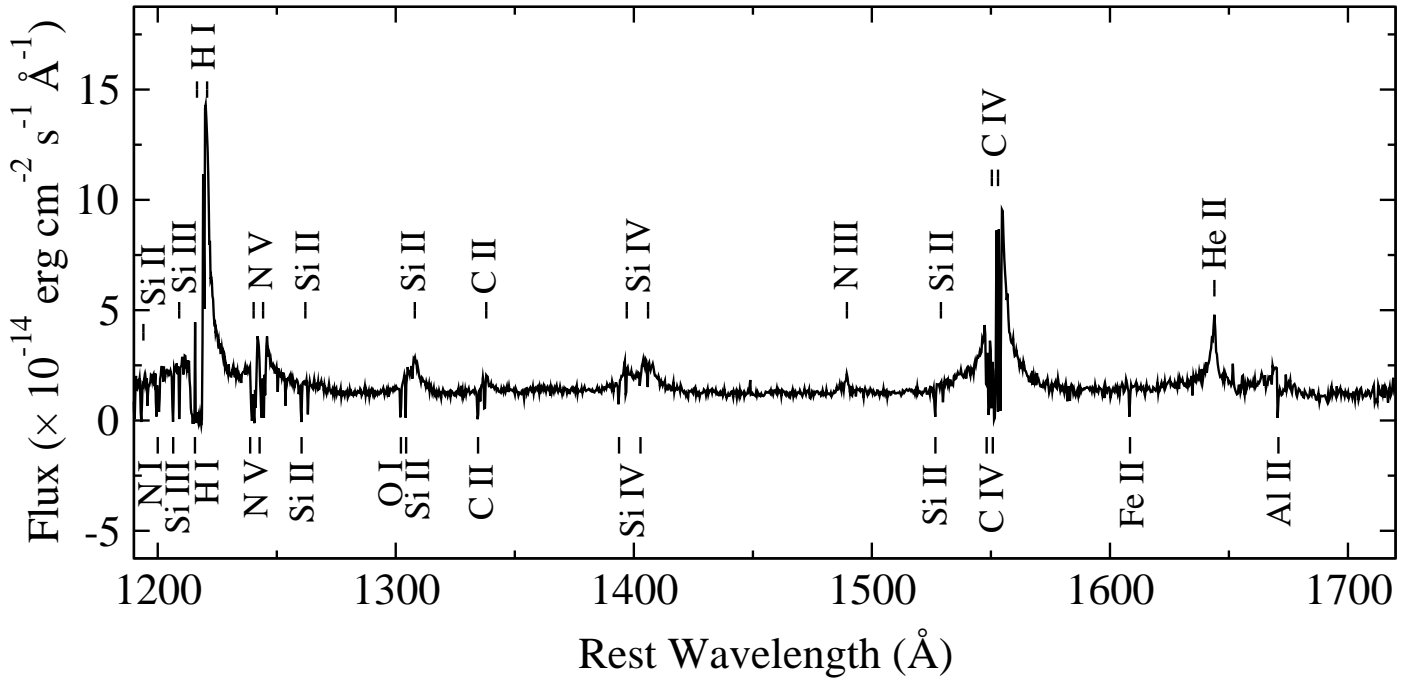}}
\caption{Heavily binned STIS spectrum of NGC\,4051. The bin size is
0.25~\AA.  Intrinsic features are labeled above the spectrum. Strong
Galactic features are labeled below the spectrum (wavelengths are from
Morton, York, \& Jenkins 1988).
\label{uvspec}}
\end{figure*}

%--------------------------------------------------------------------------------

NGC\,4051 was observed with STIS (Woodgate et~al. 1998) on {\it HST\/}
during the {\it Chandra\/} observation (see Figure~\ref{curve}).
This observation was part of the {\it HST\/} Cycle~8 guest observer
program (data sets O5F001010, O5F001020, O5F001030, and O5F001040). The
STIS observation was made with the medium-resolution echelle (E140M
mode), using an aperture size of $0\farcs2 \times 0\farcs2$. The total
exposure time was 10.3~ks (4~orbits), and the wavelength coverage was
1173--1730 \AA\ with a resolution of $\approx 7$~km~s$^{-1}$
and an absolute wavelength calibration good to within one resolution 
element. We have
analyzed the standard STScI pipeline spectrum and compared it with the
spectrum produced using the Interactive Data Language ({\sc idl})
software of the Instrument Definition Team (IDT) for STIS (see, e.g.,
Crenshaw et~al.  2000). We find good overall consistency between the
two spectra, and we use the IDT spectrum in the analysis below.
Figure~\ref{uvspec} shows the UV spectrum of NGC\,4051 from
1190--1720~\AA.

We have searched for orbit-to-orbit variability of the UV continuum,
absorption lines and emission lines. We detect none despite the
simultaneous large-amplitude X-ray variability (see Figure~\ref{curve};
compare with Done et~al. 1990). The continuum fluxes in broad
wavelength bands are consistent to within $\approx 3$\%. This lack of
detected variability justifies our use of the mean UV spectrum below.

\subsubsection{Detected Ultraviolet Absorption Lines}
\label{stobs2}

\begin{figure*}
\centerline{\includegraphics[width=14cm]{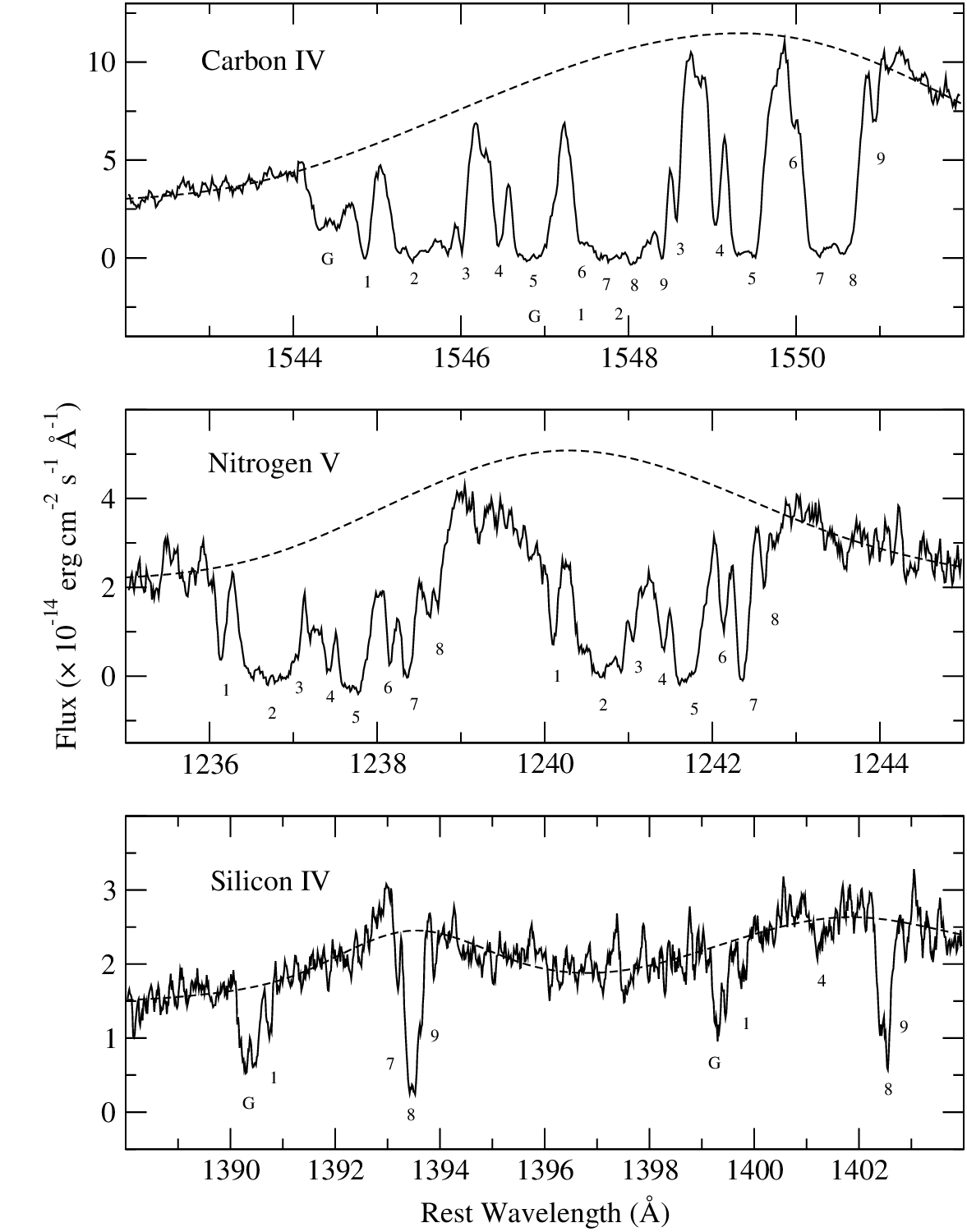}}
\caption{Intrinsic UV absorption systems in NGC\,4051. The bin sizes
for the upper, middle, and lower panels, respectively, are 0.016 \AA,
0.013 \AA, and 0.015 \AA. Spectra have been smoothed with a boxcar
filter that is five bins in width. The absorption features are marked
`G' for Galactic or numbered 1--9 for the intrinsic systems. The cubic
spline continuum fits are shown. Note that many of the absorption lines
are saturated.
\label{uvsys}}
\end{figure*}

The STIS spectrum of NGC\,4051 reveals the presence of a number of
intrinsic absorption systems with velocities ranging from
$-650$~km~s$^{-1}$ to 30~km~s$^{-1}$. The velocity widths (FWHM) of the
components range from 23~km~s$^{-1}$ to 165~km~s$^{-1}$. We see up to
nine major systems in \ion{C}{4} and \ion{N}{5}, and we detect some of
these in \ion{Si}{4}, \ion{Si}{3}, \ion{Si}{2}, and \ion{C}{2} as well
(see Table~\ref{uvlines} and Figures~\ref{uvspec}, \ref{uvsys},
and~\ref{lowsys}). We note, however, that our criteria for identifying
distinct absorption systems are somewhat subjective. We classified two
adjacent absorption systems as distinct if they appeared visually
separate in at least one of the transitions shown in
Figure~\ref{uvsys}. Unfortunately, we were unable to examine intrinsic
Ly$\alpha$ absorption from NGC\,4051, since it displays a saturated
absorption trough due to Galactic and intrinsic absorption (as well as
perhaps intergalactic material) that extends from about
$-980$~km~s$^{-1}$ to 50~km~s$^{-1}$ (in the rest frame of NGC\,4051; 
see Figure~\ref{la}).
This trough includes the entire range of observed intrinsic UV
absorption velocities and is therefore consistent with our other
results. We note that there are also two Ly$\alpha$ absorption systems
that are not saturated with velocities of $\approx 110$~km~s$^{-1}$ and
$\approx 260$~km~s$^{-1}$. These systems do not seem to have
counterparts in any of the high-ionization UV lines; they may represent
low-ionization, high-velocity clouds in NGC\,4051 (as in our Galaxy;
e.g., Wakker \& van Woerden 1997). We identify a Galactic absorption
system candidate (having velocity approximately consistent with
Galactic origin) in all ions but \ion{N}{5} (labeled `G' in
Figures~\ref{uvsys} and \ref{lowsys}), and we note that this absorption
could possibly be subdivided into two distinct systems. We also point
out that system~1 could be Galactic in origin.

%--------------------------------------------------------------------------------

%--------------------------------------------------------------------------------

In Figure~\ref{uvsys} and Table~\ref{uvlines} we present the absorption
systems from the higher ionization ions (\ion{C}{4}, \ion{N}{5}, and
\ion{Si}{4}). In order to measure the velocity shifts, FWHMs, and EWs
of the systems (many of which are saturated), we have fit the continuum
on either side of the absorption with a cubic spline. The relevant
continuum fits are shown in Figure~\ref{uvsys}. Although this fitting
technique may underestimate the continuum level if there is a
significant narrow component to the emission, visual inspection of the
less-strongly absorbed emission lines suggests that this is probably
not a highly significant effect.  Because the error on the spline-fit
continuum level is difficult to quantify, we adopt a global uncertainty
of 10\% on these values, which is consistent with the range of possible
continuum fits we considered. The errors on the equivalent widths
represent a combination of this uncertainty in the continuum level and
the actual photon noise. In general, where two or more absorption
components appear to be strongly blended together, we report the sum of
the EWs rather than attempting to separate them. The velocities 
and velocity FWHMs of the systems have been determined from
simple Gaussian fits to the absorption. The velocity shifts 
reported in Table~\ref{uvlines} represent the averages 
from \ion{N}{5}, \ion{C}{4}, and \ion{Si}{4}; we take the 
root-mean-square 
deviations of these measurements (which are typically 10--15~km~s$^{-1}$) 
to be the best estimates of the uncertainties involved.

In Figure~\ref{lowsys} we present the absorption systems from the 
lower-ionization ions (\ion{C}{2}, \ion{Si}{3}, and \ion{Si}{2}). In these
lines we identify an additional absorption system (labeled as `10')
which has a velocity of $\approx -80$~km\,s$^{-1}$. 
For the lower-ionization ions, we tentatively identify counterparts to 
some of the absorption systems found using the higher-ionization ions
(the statistical significances of these features are strongly dependent 
upon the adopted continuum level).
The most notable counterparts are
systems~1, 4, 7, 8, and 9, which are found in all ions,
system 2 (found in \ion{Si}{2} and \ion{Si}{3}), and 
system 5 (found in \ion{C}{2} and \ion{Si}{3}).
The low-velocity absorption systems (8--10) may originate within the
host galaxy rather than the active nucleus.  However, the high-velocity
absorption systems (2--5) seen in the low-ionization ions are most
likely related to the nucleus.  Such high-velocity absorption systems
in low-ionization ions have been observed in only one other Seyfert
galaxy, NGC\,4151 (Weymann et~al. 1997).

\subsubsection{Comparison with {\it IUE} and Ultraviolet Emission Lines}
\label{stobs3}

Peterson et al. (2000) present \ion{He}{2} and \ion{C}{4} emission
lines for NGC\,4051 from a mean UV spectrum of 31 observations carried
out \ by \ the \ {\it \ International \ Ultraviolet \ Explorer \ (IUE)}

%--------------------------------------------------------------------------------

\centerline{}
%\begin{figure*}
\centerline{\includegraphics[width=8.7cm]{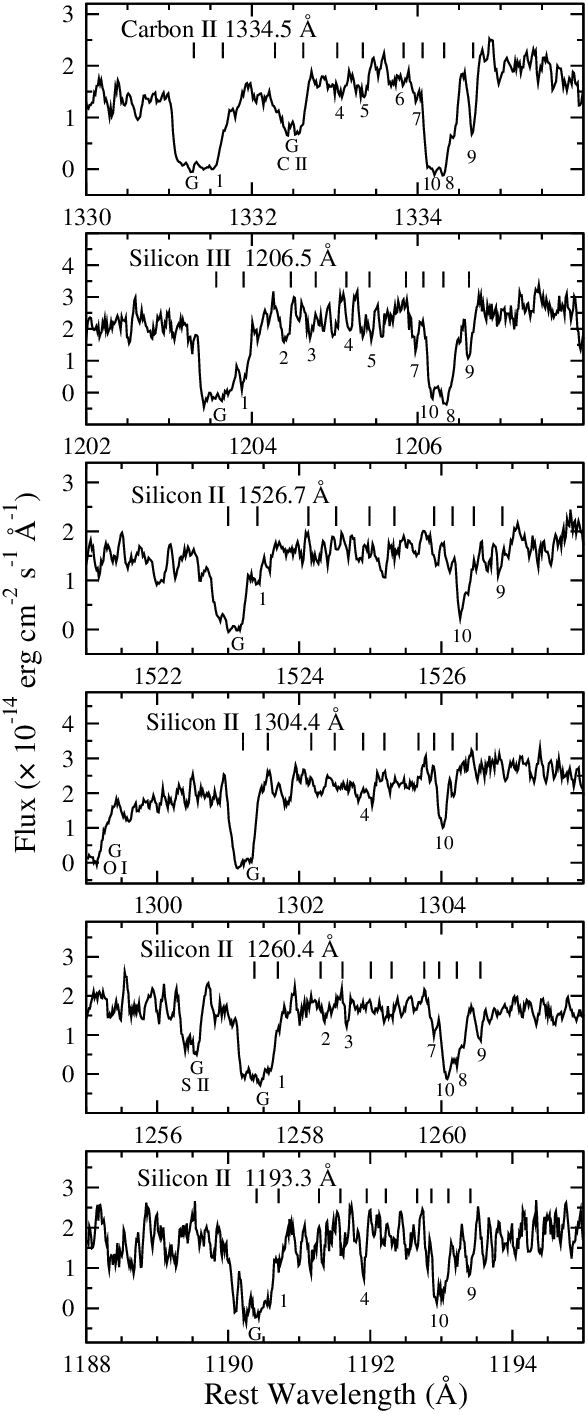}}
\figcaption{Intrinsic UV absorption systems in NGC\,4051 from the
low-ionization ions \ion{C}{2}, \ion{Si}{3}, and \ion{Si}{2}. The bin
size is $\approx 0.15$~\AA. Spectra have been smoothed with a boxcar
filter that is five bins in width. The expected wavelengths of the
absorption systems from Figure~\protect\ref{uvsys} and
Table~\protect\ref{uvlines} are
marked as vertical lines in each panel. Some detected absorption
features are labeled as in Figure~\protect\ref{uvsys}. Here we also identify an
additional absorption system, labeled as `10', which has a blueshifted
velocity of $\approx -80$~km\,s$^{-1}$.
\label{lowsys}}
%\end{figure*}
\centerline{}

%--------------------------------------------------------------------------------
%--------------------------------------------------------------------------------

%\begin{figure*}
\centerline{\includegraphics[width=8.5cm]{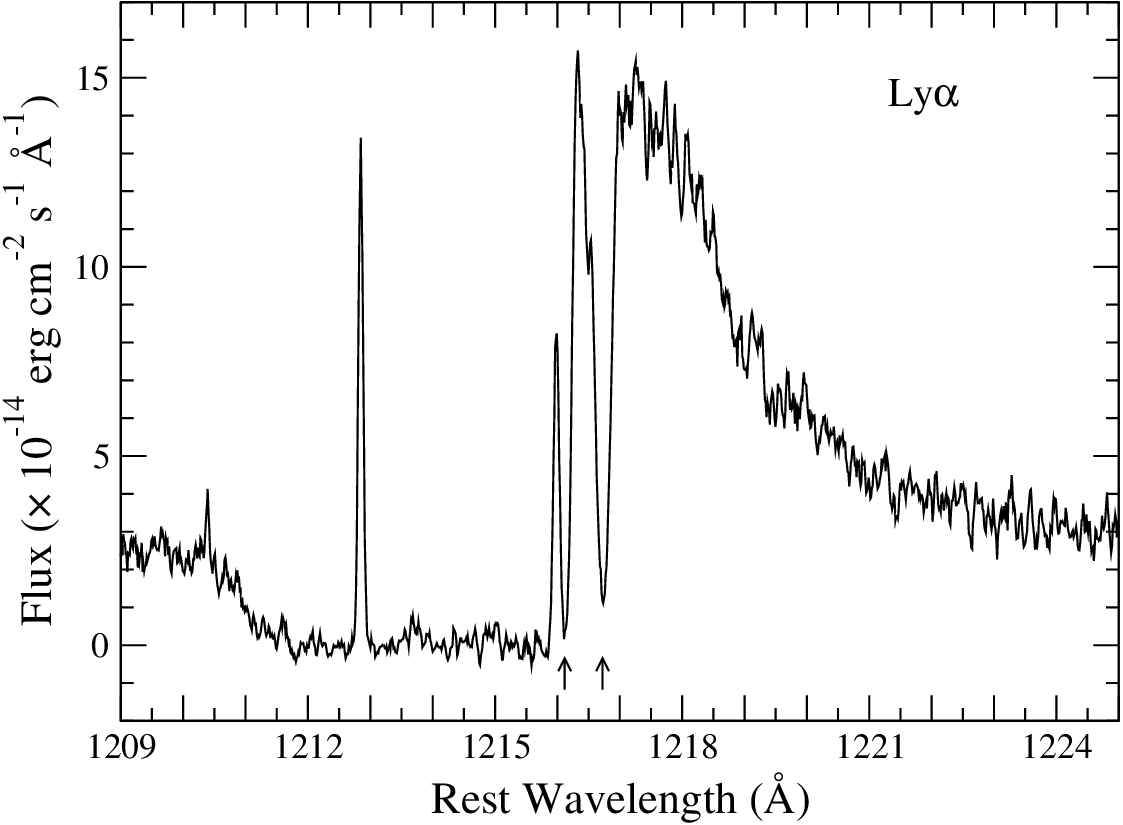}}
\figcaption{Absorption near Ly$\alpha$ in the spectrum of NGC\,4051. The
bin size is $0.13$~\AA. The spectrum has been smoothed with a boxcar
filter that is five bins in width. Note the wide absorption trough and
the narrow geocoronal emission line near 1213~\AA\ (the spectrum has 
been corrected for the cosmological redshift). The two absorption systems
with redshifted velocities of $\approx 110$~km~s$^{-1}$ and 
$\approx 260$~km~s$^{-1}$ are marked with arrows. These systems may 
originate in low-ionization, high-velocity clouds within NGC\,4051 
(as in our Galaxy).
\label{la}}
%\end{figure*}
\centerline{}
\centerline{}

%--------------------------------------------------------------------------------

%--------------------------------------------------------------------------------

%\begin{figure*}
\centerline{\includegraphics[width=8.5cm]{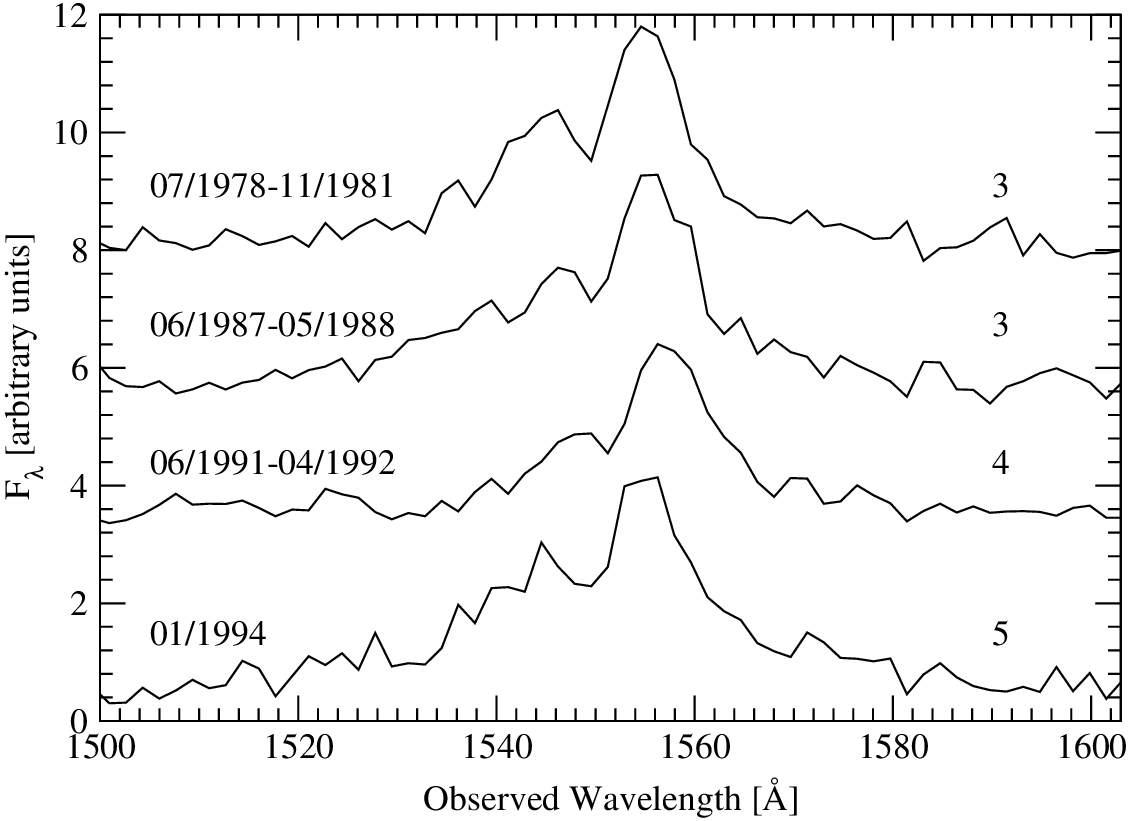}}
\figcaption{Mean {\it IUE} spectra of NGC\,4051 (around the \ion{C}{4}
emission line) for several epochs. The epochs are presented on the left,
and the number of spectra averaged is on the right. Note the absorption
$\approx 2.5$~\AA\ blueward of the line peak, which is present in all
epochs and is consistent with the absorption seen in the {\it HST STIS}
spectrum. The small wavelength shifts between the spectra are probably
due to wavelength calibration problems and the low resolution. These
small wavelength shifts are one reason that the absorption is difficult
to see in the mean spectrum presented by Peterson et~al. (2000).
\label{new_uv_fig}}
%\end{figure*}
\centerline{}
\centerline{}

%--------------------------------------------------------------------------------

\begin{figure*}
\centerline{\includegraphics[width=16cm]{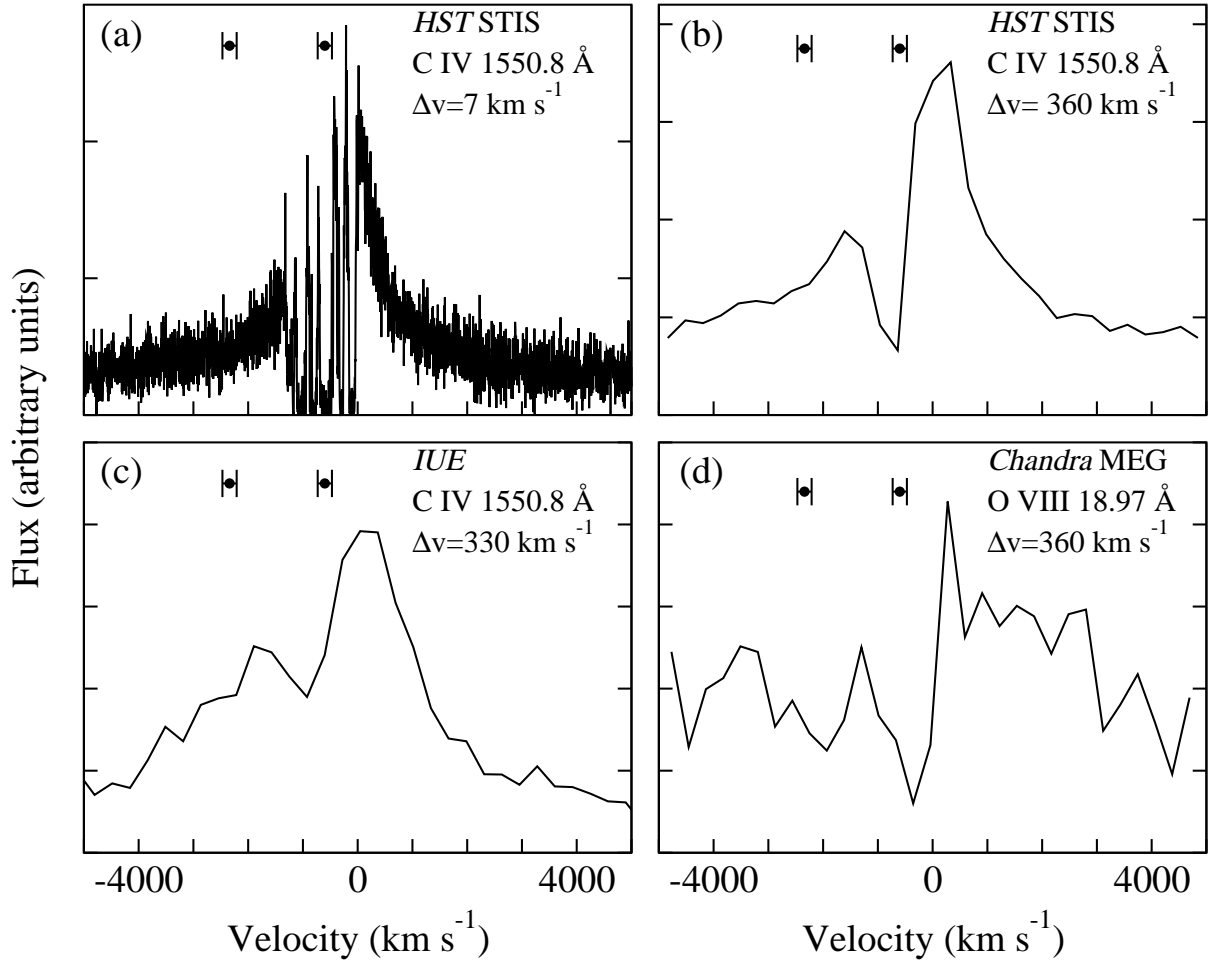}}
\caption{Absorption and emission systems as seen by (a) and (b) STIS,
(c) {\it IUE} (mean of the 1978--1988 and 1994 epochs presented in
Figure~\ref{new_uv_fig}), and (d) {\it Chandra}; panel (b) represents
the STIS \ion{C}{4} line binned to the velocity resolution of the MEG.
The approximate velocity resolution is given for each plot.  The
velocity units in the panels showing \ion{C}{4} correspond to the
1550.8~\AA\ line, and dots with error bars mark the velocities of the
two X-ray absorption components resolved by {\it Chandra\/}.
Comparison of the panels stresses the point that the X-ray warm
absorber may be subdivided into further systems that cannot be resolved
with current X-ray instruments.  Note also that no strong ultraviolet
absorption is seen corresponding to the high-velocity X-ray absorption
component.
\label{rescomp}}
\end{figure*}

\noindent Short-Wavelength Prime camera (SWP; wavelength range
$\sim$\,1150--2000~\AA) over a period of 16 years.  Though it is not
clear from Peterson et al. (2000), a careful plotting of the \ion{C}{4}
line reveals the likely existence of absorption in 90\% of the spectra
(see Figures~\ref{new_uv_fig} and~\ref{rescomp}; the lack of apparent
absorption in the other spectra is probably due to their low S/N). Thus
it appears that \ion{C}{4} absorption has existed in NGC\,4051 since at
least 1978.

%--------------------------------------------------------------------------------

%--------------------------------------------------------------------------------

The {\it STIS} spectrum of NGC\,4051 is consistent with the mean {\it
IUE} spectrum at wavelengths longer than 1225~\AA. Around the
Ly$\alpha$ line the {\it IUE} spectrum is contaminated by geocoronal
Ly$\alpha$ emission, and no comparison is possible. The emission lines
from both spectra agree well in their profiles. The blueward asymmetry
of the \ion{He}{2} and \ion{C}{4} broad emission described by Peterson
et~al. (2000) is also detectable in the {\it STIS} spectrum. This
effect is also seen in other NLS1s (e.g., Leighly 2000). We list the 
UV emission-line fluxes in Table~3; in all cases the 
fluxes have been measured from unabsorbed line profiles fit to the 
data as described in \S~\ref{stobs2}. Note that the 
fluxes given for Ly$\alpha$, \ion{N}{5}, and \ion{C}{4} are quite 
uncertain due to their heavily absorbed profiles.

Although the strong absorption and blueward asymmetry make measurements
of the widths of the UV emission lines difficult, we have attempted to
measure the FWHMs of the \ion{C}{4} 1548.2~\AA\ and 1550.8~\AA\ lines
from the STIS data.  To do this, we started by assuming the centers of
these lines are consistent with rest-frame emission in NGC\,4051.  We
have attempted to deblend the features by fitting the cubic spline
representing the continuum in this area with two Gaussians centered at
the fixed wavelengths of the transitions; we required the relative strengths 
of the 1548.2~\AA\ and 1550.7~\AA\ lines to lie between 1:1 and 2:1. 
After obtaining the best
fit, we blueshifted the lines by steps of 50~km\,s$^{-1}$ and repeated
the fitting. We find the best fit to be when the lines are blueshifted
by $\approx 100$~km~s$^{-1}$ and the strengths of the lines are approximately 
equal. The FWHMs of the lines are $\approx 1040\pm250$~km~s$^{-1}$. 
These FWHMs, although rather uncertain, are
consistent with the value of $1110\pm 190$~km~s$^{-1}$ reported by
Peterson et~al. (2000) for the FWHM of the variable broad component of
H$\beta$.

%--------------------------------------------------------------------------------

\section{DISCUSSION}

\subsection{X-ray and UV Absorption and Emission}
\label{xuv}

The most significant results of this analysis are the discovery of
X-ray emission lines and blueshifted X-ray absorption lines similar to
those recently found from several other Seyfert~1s (e.g., Kaastra
et~al. 2000; Kaspi et~al. 2000), as well as the identification of
multiple intrinsic X-ray and UV absorption systems in the spectrum of
NGC\,4051. The velocities of the X-ray absorption components are
measured to be $-2340\pm 130$~km~s$^{-1}$ (the X-ray warm absorber with
the highest velocity yet discovered) and $-600\pm 130$~km~s$^{-1}$,
with an average velocity for the emission lines (excluding
Fe~K$\alpha$) of $14\pm 120$~km~s$^{-1}$.  The low-velocity X-ray
absorption system, taking into account its measured width of 
$750^{+400}_{-280}$~km~s$^{-1}$, is
\ consistent \ with \ the \ velocity

%--------------------------------------------------------------------------------

\end{multicols}
\begin{deluxetable}{lcccccccc}
\tabletypesize{\footnotesize}
\tablecolumns{9}
\tablewidth{0pt}
\tablecaption{\sc UV Absorption Systems in the {\it HST} STIS Spectrum
\label{uvlines}}
\tablehead{
\colhead{} &
\colhead{} &
\colhead{} &
\colhead{N V} &
\colhead{N V} &
\colhead{Si IV} &
\colhead{Si IV} &
\colhead{C IV} &
\colhead{C IV} \\
\colhead{} &
\colhead{} &
\colhead{Velocity} &
\colhead{$\lambda$ 1238.8} &
\colhead{$\lambda$ 1242.8} &
\colhead{$\lambda$ 1393.8} &
\colhead{$\lambda$ 1402.8} &
\colhead{$\lambda$ 1548.2} &
\colhead{$\lambda$ 1550.8} \\
\colhead{Sys.} &
\colhead{Velocity} &
\colhead{FWHM} &
\colhead{EW} &
\colhead{EW} &
\colhead{EW} &
\colhead{EW} &
\colhead{EW} &
\colhead{EW} \\
\colhead{No.} &
\colhead{(km s$^{-1}$)} &
\colhead{(km s$^{-1}$)} &
\colhead{(m\AA)} &
\colhead{(m\AA)} &
\colhead{(m\AA)} &
\colhead{(m\AA)} &
\colhead{(m\AA)} &
\colhead{(m\AA)}}
\startdata

G & $-727$  & $\phn 74$ & 
\nodata  & \nodata  & 
$240^{+32}_{-29}$ & $130^{+37}_{-32}$ & 
$301^{+31}_{-26}$  & $537^{+10}_{-9}$\tablenotemark{a} \\

1 & $-647$  & $\phn 40$ & 
$110^{+14}_{-13}$  & $125^{+7}_{-6\phn}$  & 
$\phn54^{+16}_{-14}$ & $\phn40^{+26}_{-23}$ & 
$223^{+12}_{-11}$  & $253^{+9}_{-8}$\tablenotemark{a} \\

2 & $-505$  & $ 165$ & 
$\phn628\pm 15$  & $595\pm 8\phn$  & 
\nodata & \nodata & 
$750^{+17}_{-15}$ & $710\pm 4$\tablenotemark{a}  \\

3 & $-430$  & $\phn 63$  & 
$123\pm 7$  & $109^{+6}_{-5\phn}$  & 
\nodata & \nodata & $133\pm 6$ & $100^{+7}_{-6\phn}$  \\

4 & $-337$  & $\phn 52$  & 
$121^{+5}_{-4}$  & $115^{+5}_{-4\phn}$  & 
\nodata & $\phn 24^{+15}_{-13\phn}$ & $168^{+9}_{-8\phn}$ & $121^{+8}_{-7\phn}$  \\

5 & $-268$  & $ 133$ & 
$425\pm 7$  & $441^{+13}_{-12}$  & 
\nodata & \nodata & 
$537^{+10}_{-9}$\tablenotemark{a} & $418^{+7}_{-6\phn}$   \\

6 & $-158$  & $\phn 45$  & 
$121\pm 5$  & $104^{+9}_{-8\phn}$  & 
\nodata & \nodata & 
$253^{+9}_{-8}$\tablenotemark{a} & $\phn39^{+10}_{-8}$   \\

7 & $-107$ & $\phn 64$  & 
$179\pm 5$  & $196^{+11}_{-10}$  & 
$\phn30^{+12}_{-10}$ & \nodata & 
$710\pm 4$\tablenotemark{a} & $360^{+8}_{-7\phn}$   \\

8 & $\,\,-48$   & $\phn 84$   & 
$154^{+14}_{-12}$  & $\phn64^{+23}_{-19}$  & 
$281^{+21}_{-18}$ & $189^{+30}_{-26}$ & 
$710\pm 4$\tablenotemark{a} & $300^{+9}_{-7\phn}$   \\

9 & $\phm{-1}30$    & $\phn 23$  & 
\nodata & \nodata & 
$\phn22^{+9\phn}_{-8}$ & $\phn16^{+11}_{-9}$ & 
$152\pm 3$\tablenotemark{b} & $\phn31^{+11}_{-9}$  \\
\enddata
%\vskip 0.2 in
\tablecomments{The FWHM values represent the averages from the N~V lines, 
except for systems G and 9 which are from \ion{C}{4} and \ion{Si}{4}. The 
uncertainties on 
the system velocities and FWHMs are roughly 10--15~km~s$^{-1}$. EW errors are 
reported at the 1$\sigma$ level.}
\tablenotetext{a}{Blended with one or more other components (see 
Figure~\ref{uvsys} for clarification). The components that are blended too 
strongly to separate their EWs are the following: 
component~5 (1548~\AA) and component~G 
(1551~\AA); component~6 (1548~\AA) and component~1 (1551~\AA); components~7 
and~8 (1548~\AA) and component~2 (1551~\AA).} 
\tablenotetext{b}{Component~9 (1548~\AA) is also strongly blended 
with components~7 and~8 (1548~\AA) and component~2 (1551~\AA).}
\end{deluxetable}
\begin{multicols}{2}

%--------------------------------------------------------------------------------

\noindent range of the intrinsic UV absorption
systems from $-650$~km~s$^{-1}$ to $-270$~km~s$^{-1}$ (systems~1--5).
On the other hand, the high-velocity X-ray absorption system appears to
have no kinematically consistent counterpart in the UV (see
Figure~\ref{rescomp}a), and the UV absorption systems with the smallest
blueshifts appear to have no strong counterparts in the X-rays. These
observations are physically consistent, however, because the
lowest-velocity UV absorption systems are seen preferentially in
low-ionization species while the high-velocity X-ray absorption system
is seen only in very high-ionization species. Another notable point is
that the X-ray absorption may well be subdivided into more than the two
systems we are able to resolve with the HETGS. For example,
Figure~\ref{rescomp} shows the gain in information between {\it IUE}
resolution and STIS resolution; there may well be equal complexity awaiting
discovery in X-rays. This warning should be remembered in efforts to
perform detailed modeling of the X-ray absorption lines, particularly
when using data from the {\it XMM-Newton\/} Reflection Grating
Spectrometer (RGS) which has lower resolution than the {\it Chandra\/}
HETGS.

%--------------------------------------------------------------------------------

\begin{figure*}
\centerline{\includegraphics[width=16cm]{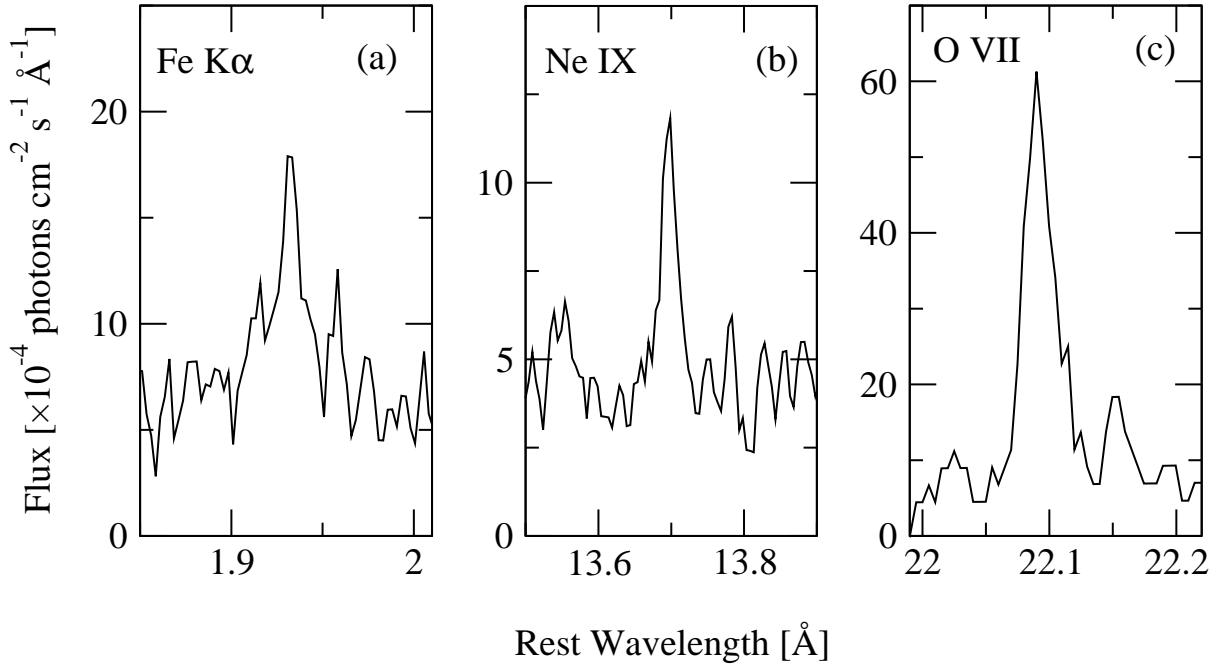}}
\caption{Strongest X-ray emission lines from NGC\,4051. Panel (a)
is from the HEG first-order spectrum, while panels (b) and (c) are from
the MEG first-order spectrum. All features have been smoothed using a
boxcar filter that is three bins in width. Note the differences in
vertical scale among the panels.
\label{emission}}
\end{figure*}

%--------------------------------------------------------------------------------

In terms of ionization potential, the high-velocity X-ray absorption
component is observed preferentially in the higher-ionization (H-like)
ions, while the low-velocity component is stronger in the
lower-ionization (He-like) ions (i.e., the ionization potentials to 
{\it create} the H-like ions in Figure~\ref{profiles} 
from their He-like counterparts are all higher than the
ionization potentials to {\it create} the He-like ions in this figure 
from their Li-like counterparts), although its presence is 
also statistically suggested in some of the H-like ions. This
ionization potential approach was phenomenologically motivated by the
observed similarities between many of the velocity spectra for the 
H-like ions as a group and the He-like ions as a group.
However, it is also relevant to compare the ionization parameters for
which the relative populations of particular ions are maximized
(assuming photoionization equilibrium). Kallman \& McCray (1982)
present curves showing the relative fractions of different ionization
states of O, Ne and Si as functions of ionization parameter in the
context of several different X-ray nebular models.  Examination of
these curves reveals that the population of He-like \ion{Si}{13} is
expected to peak at a larger ionization parameter than either of the
H-like ions \ion{O}{8} and \ion{Ne}{10}, which seems to conflict with
the ionization potential approach used to construct
Figure~\ref{profiles}. Detailed modeling will be required to ascertain
if there is truly a conflict, but unfortunately such modeling will
prove difficult if the absorbing gas in NGC\,4051 is not in
photoionization equilibrium (e.g., Krolik \& Kriss (1995); 
McHardy et~al. 1995; Nicastro et~al. 1999).
NGC\,4051 typically shows large-amplitude variability of its average
flux on timescales ranging from $\sim 100$~s to a couple days (see 
Figure~\ref{curve} and 
McHardy, Papadakis, \& Uttley 1998), so if the recombination timescale
of the warm absorber lies in this range it is unlikely to be in
photoionization equilibrium. All states of highly ionized oxygen, for
example, recombine on a timescale given by $t_{\rm rec} \approx 200
n_9^{-1}T_5^{0.7}$~s, where $n=10^9n_9$~cm$^{-3}$ and $T=10^5T_5$~K
(Shull \& Van~Steenberg 1982).  Adopting a typical warm absorber
temperature of $2\times 10^5$~K, we find that the warm absorber will
fall out of photoionization equilibrium if its density lies between
$\sim 10^6$~cm$^{-3}$ and $\sim 3\times 10^9$~cm$^{-3}$.\footnote{If
the density is $\ll 10^6$~cm$^{-3}$, corresponding to a value of
$t_{\rm rec}$ much longer than the typical variability timescale, then
the warm absorber might be in equilibrium with the {\it average\/}
continuum flux; it would not respond to short-timescale continuum flux
variations.} This fairly wide range of densities is consistent with the
constraints derived using He-like lines and with 
the upper end of the range of plausible densities for the Narrow Line
Region (see below for the He-like line constraints and dynamical
consistency between the Narrow Line Region lines and the X-ray warm
absorber lines).
Hence the warm absorber of NGC\,4051 is plausibly not in
photoionization equilibrium, and our phenomenological approach based on
ionization potential seems appropriate for this analysis. The limited
photon statistics prevent us from investigating the details of the
ionization physics via timing analyses (e.g., Nicastro et~al. 1999).

The strongest emission features seen in the X-ray spectrum of NGC\,4051
are the \ion{O}{7} 22.1~\AA\ and \ion{Ne}{9} 13.7~\AA\ lines. Such
features have been predicted to arise from the X-ray absorbing gas
(e.g., Krolik \& Kriss 1995; Netzer 1996) and have now been seen from
other Seyfert~1s as well. The low Galactic column density to NGC\,4051
allows for a full appreciation of the strengths of these lines, which
are shown (along with Fe~K$\alpha$) in Figure~\ref{emission}. The
best-fit FWHMs of these lines are $357^{+208}_{-220}$~km~s$^{-1}$ and
$554^{+396}_{-320}$~km~s$^{-1}$, respectively, which are consistent
with being unresolved by the MEG. The upper limit on the width of the 
\ion{O}{7} line is significantly below the observed widths of the
broad optical and ultraviolet lines. If a simple anticorrelation of
line width and radial location in a virialized system is appropriate,
this line must arise in a region farther from the central source
than the Broad Line Region. It may arise in the Narrow Line Region,
which for NGC\,4051 has a characteristic [\ion{O}{3}] FWHM of
210--330~km~s$^{-1}$ (e.g., Heckman et~al. 1981; De Robertis \&
Osterbrock 1984). However, we note that the \ion{O}{7} line probably 
arises in a higher-ionization gas component than the [\ion{O}{3}] 
line. An alternative explanation is that this 
high-ionization line may originate in the gas that produces the optical 
coronal lines (e.g., Porquet et~al. 1999).

The Fe~K$\alpha$ line from NGC~4051 (see Figure~\ref{emission} and 
Table~2) also deserves mention. The upper limit on the 
width of the {\it Chandra\/} 
line (FWHM) is $\approx 2800$~km~s$^{-1}$, substantially 
below the width of the broad accretion-disk line observed by {\it ASCA} 
(e.g., G96). The line is also consistent with being unresolved by the 
HEG. The line's width, along with its measured EW of $158^{+51}_{-47}$~eV 
($47.7^{+15.4}_{-14.2}$~m\AA) and energy of $6.41^{+0.01}_{-0.01}$~keV, 
are consistent with an origin 
in neutral material surrounding the nuclear region, such as the 
putative molecular torus of AGN unification schemes (e.g., Krolik, Madau, 
\& Zycki 1994). We note that the EW of this relatively 
narrow feature is a significant fraction of the EW of the 
line seen by G96 ($\approx 300$~eV). This point should 
be remembered in future attempts to model the broad Fe~K$\alpha$ line.

Since the UV absorption seen from NGC\,4051 strongly absorbs the broad
UV emission-line flux, it is clear that the gas in which this
absorption arises must be coincident with or farther from the central
engine than the UV Broad Line Region. Our estimate of the width of the
\ion{C}{4} lines, which is consistent with the width of the H$\beta$
line, suggests that the physical sizes of the optical and UV Broad Line
Regions are roughly comparable.  This sets a rough lower limit of 3--6~light
days (Peterson et~al. 2000) on the distance from the central source to
the UV-absorbing gas.

\subsection{Column Density and Density Estimates}

We have estimated ionic column densities using the X-ray
absorption-line EWs and the `curve of growth' technique described in
\S~3.4 of Spitzer (1978). In these analyses, we have treated the
high-velocity and low-velocity X-ray absorption systems separately.
Unfortunately, the column densities for the high-velocity system
are quite uncertain because 
(1) the lines are not resolved so we can only constrain the 
velocity-spread parameter $b$ to be $\la 1000$~km~s$^{-1}$ and  
(2) we are near or on the `flat' part of the curve of growth. 
Furthermore, if the X-ray absorption lines have unresolved substructure
(see Figure~\ref{rescomp}) this may confuse curve of growth analyses.
The two X-ray absorption lines in the high-velocity system
with the best photon statistics are \ion{Mg}{12} 8.419~\AA\ and 
\ion{Si}{14} 6.181~\AA. If we require $b\ga 100$~km~s$^{-1}$,
for example, the \ion{Mg}{12} column density formally allowed by 
our curve of growth analysis is between 
$\approx 7\times 10^{16}$~cm$^{-2}$ and 
$\approx 4\times 10^{20}$~cm$^{-2}$. 
We consider the upper end of this range to be implausibly large (e.g.,
the implied absorber becomes optically thick to electron scattering),
but the lower end is plausible.  For solar abundances and the likely
ionization correction, the corresponding lower bound on $N_{\rm H}$ is
$\sim 2\times 10^{21}$~cm$^{-2}$.  Similarly, for \ion{Si}{14} the
column density formally allowed is between 
$\approx 9\times 10^{16}$~cm$^{-2}$ and 
$\approx 5\times 10^{20}$~cm$^{-2}$. 
Assuming solar abundances, the column densities of these two ions
could plausibly arise in a zone characterized by a single ionization 
parameter, although this is not surprising given the large uncertainties.
We note that the \ion{O}{8} column density of 
$(1.4\pm 0.3)\times 10^{18}$~cm$^{-2}$ derived from the absorption edge depth 
is consistent with the plausible range of column densities in the high-velocity 
X-ray absorption line system.

The ionic column densities of the low-velocity system are easier
to constrain because 
(1) the \ion{O}{7} lines are resolved and thus we can measure $b$ directly and 
(2) we are on the linear part of the curve of growth. The
low-velocity lines most useful for this analysis are \ion{O}{7}
17.768~\AA, 18.628~\AA\ and 21.602~\AA. The allowed \ion{O}{7} column
density ranges for these lines are, respectively,
(5--20)$\times 10^{16}$~cm$^{-2}$, 
(3--7)$\times 10^{16}$~cm$^{-2}$, and 
(8--20)$\times 10^{15}$~cm$^{-2}$.
Although these ranges are not all consistent with a single value, they
are not wildly different. With solar abundances and a plausible
ionization correction, the implied $N_{\rm H}$ is between $\sim
1\times 10^{19}$~cm$^{-2}$ and $\sim 5\times 10^{20}$~cm$^{-2}$. 
We point out, however, that the \ion{O}{7} 
column density of $(8\pm 1)\times 10^{17}$~cm$^{-2}$ indicated 
from the absorption-edge depth is larger than the upper limit derived from 
the curve of growth technique. The seeming inconsistencies among the 
implied column densities may result from either the effects of iron L-shell
absorption lines near the \ion{O}{7} absorption edge or the existence of 
multiple unresolved absorption components within the low-velocity X-ray 
absorption system.

We note that the optical
spectrum of NGC\,4051 contains a small contribution from polarized
light (e.g., Grupe et~al. 1998). If there is a corresponding scattered
X-ray continuum component (e.g., if the flux is due to Thompson scattering), 
non-zero flux levels may be observed in the
troughs of absorption lines that are saturated along the primary line
of sight. An effect of this type could confuse column density estimates made 
using the techniques detailed above. However, because the polarization 
fraction is small and
because no noticeable effect of this type is observed for the saturated
UV absorption lines (see \S~\ref{stobs1}), this effect should hopefully be
small for the X-ray lines as well.

The emission lines from the He-like ions we identified can serve as
plasma density diagnostics (e.g., Porquet \& Dubau 2000). For both \ion{O}{7}
and \ion{Ne}{9} we identified, in emission, only the intercombination
($i$) and forbidden ($f$) lines and not the resonance ($r$) line. The
forbidden-to-intercombination line ratios are $f/i=4.8^{+2.1}_{-6.5}$,
and $5.0^{+2.5}_{-10.2}$ for \ion{O}{7} and \ion{Ne}{9}, respectively
(errors are estimated by the numerical method described in \S~1.7.3 of
Lyons 1991). As the resonance emission line is not identified, its flux
is lower than the intercombination line flux, and the forbidden-to-resonance 
line ratio ($f/r$) has to be $\ga4$. Such high
$f/r$ implies that the plasma emitting the lines is photoionization
dominated (with little collisional ionization), and the upper limit on
its temperature is $10^{6}$~K. The upper limit on the plasma density,
using the $f/i$ ratios and Figure~8 of Porquet \& Dubau
(2000), is then $4\times 10^{10}$ cm$^{-3}$.

The S/N of the X-ray spectrum of NGC\,4051 is not favorable for
detailed photoionization modeling such as Kaspi et~al. (2001) have
performed for NGC\,3783. However, as Figure~\ref{ew_comp} shows, the
correlation between the EWs of spectral features from NGC\,4051 and
from NGC\,3783 is striking (although NGC\,3783 has stronger X-ray
absorption overall).  This suggests that the basic physical nature of
the absorbing gas is sim-
 
%--------------------------------------------------------------------------------

%\begin{figure*}
\centerline{\includegraphics[width=8.5cm]{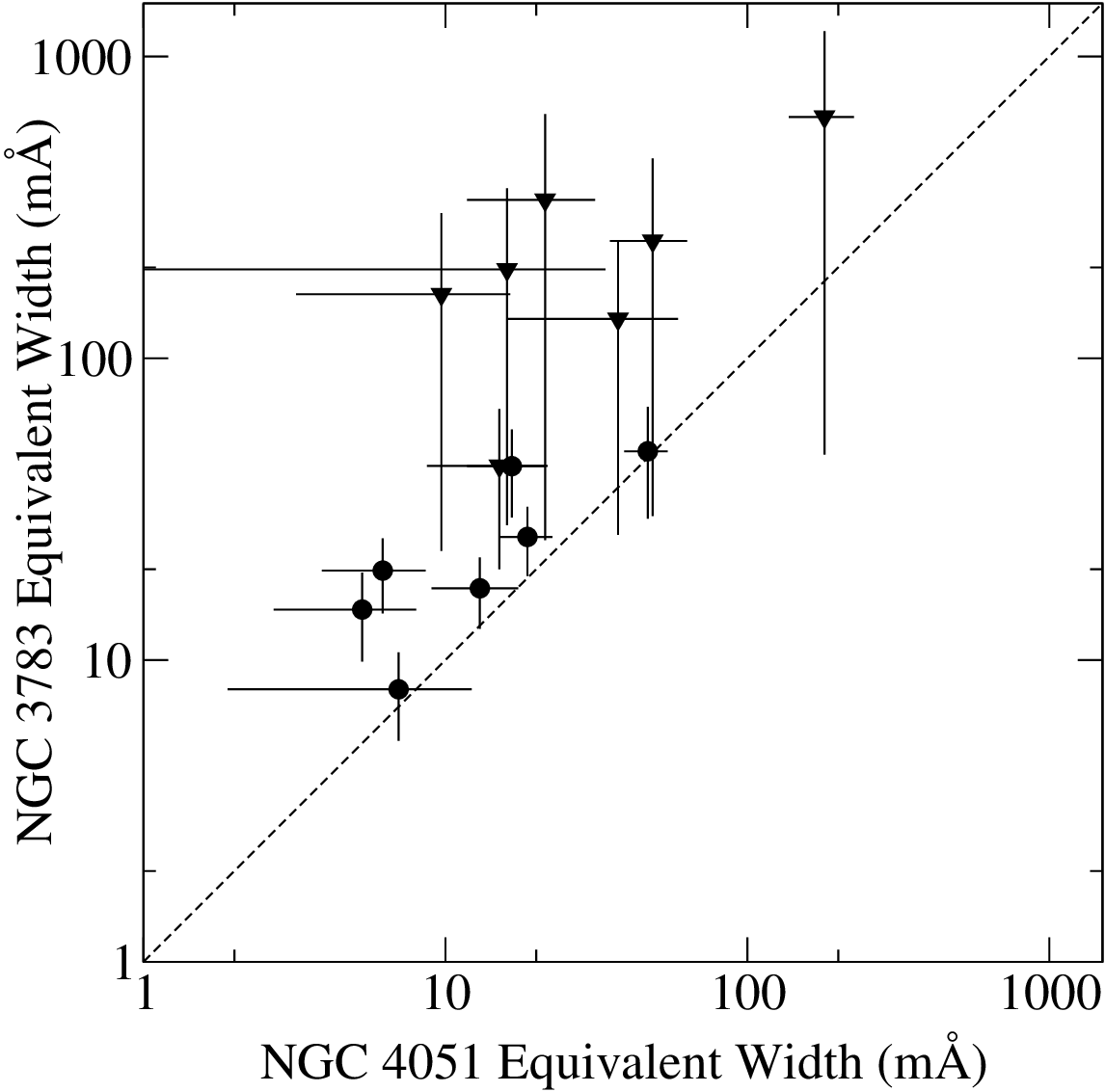}}
\figcaption{Comparison of the EWs of features detected in the {\it
Chandra\/} spectra of both NGC\,4051 and NGC\,3783 (Kaspi et~al. 2000).
Dots represent absorption features, and triangles represent emission
features. EWs of absorption lines from NGC\,4051 that are detected in both 
the high-velocity and low-velocity systems represent the sums of the 
measured values for the two systems. 
A line with slope equal to unity is plotted to guide the eye. 
Although the features seen from NGC\,4051 are generally
weaker than those seen from NGC\,3783, the noticeable correlation
spanning a factor of $\approx 100$ in EW indicates a basic similarity
in the warm absorbers of the two systems.
\label{ew_comp}}
%\end{figure*}
\centerline{}
\centerline{}

%--------------------------------------------------------------------------------

\noindent ilar in these two Seyferts and that the results
of the detailed modeling of the absorber in NGC\,3783 should be
applicable, at least to first order, for NGC\,4051 as well.  This
correlation is notable given (1) the different 0.5--10~keV ionizing
continuum shapes of the two objects (while their average hard power-law
slopes are similar, NGC\,3783 lacks the strong soft X-ray excess seen from
the NLS1 NGC\,4051) and (2) the fact that NGC\,4051 is $\approx 100$ times less
luminous in X-rays than NGC\,3783.
However, it is also important to note that the UV absorption in
NGC\,4051 is much stronger than in NGC\,3783 (while the opposite is
true for the X-ray absorption). Crenshaw et~al. (1999) report EWs of
$0.28\pm 0.04$~\AA\ and $0.19\pm 0.03$~\AA\ for the \ion{N}{5}
1238.8~\AA\ and 1242.8~\AA\ lines from NGC\,3783, while the total
values we measure for NGC\,4051 are roughly seven and nine times
larger, respectively. The combination of weaker X-ray absorption and stronger 
UV absorption seems to indicate that the absorbing gas in NGC\,4051 is 
characterized by a smaller ionization parameter than the gas in NGC\,3783.
This may result from the different
extreme-ultraviolet to soft X-ray continuum shapes of the two objects
(since this is the ionizing continuum relevant for the UV lines) or
from the difference in luminosity between the two objects; 
observations of additional
Seyferts are required to search for connections of these types.

%--------------------------------------------------------------------------------

\subsection{Variability and Soft X-ray Excess}
\label{varsoft}

The low state that NGC\,4051 entered for the last $\approx 15$~ks of our 
{\it Chandra\/} observation is comparable to the lowest states 
in which it has been previously observed, although it is not the lowest. Guainazzi 
et~al. (1998) reported a state in which the average 2--10~keV flux was 
$1.3\times 10^{-12}$~erg~cm~$^{-2}$~s$^{-1}$, smaller by a factor of about 
four than that which we observe. The flux during our low state is comparable 
to that seen during the 1990 {\it Broad-Band X-ray Telescope (BBXRT)\/} 
low-state observation (Weaver 1993). Previous analyses have found that 
the 2--10~keV spectrum of NGC\,4051 becomes increasingly hard with decreasing 
flux (e.g., Matsuoka et~al. 1990; G96; Uttley et~al. 1998). Our 
data show the expected qualitative behavior (see \S2.1.3), 
and quantitatively our hard photon index of
$\Gamma=0.83^{+0.18}_{-0.19}$ for the low state is consistent with the 
$\Gamma=0.8\pm 0.3$ observed by {\it BBXRT\/} (as well as the 
$\Gamma=0.78^{+0.37}_{-0.13}$ observed by Guainazzi et~al. 1998). 
However, both of these values lie 
somewhat below a linear extrapolation of the photon index 
versus X-ray flux correlation established for NGC\,4051 at higher
fluxes; either this correlation becomes nonlinear at low fluxes
or we are witnessing the effects of its long-term secular evolution. 
The flat photon index is probably due to a large contribution
from a reflection component that lags the primary continuum variations, 
but unfortunately our $\approx 15$~ks of low-state {\it Chandra\/} data 
lack the photon statistics to examine this matter in detail. 

Our high-resolution data on the soft X-ray excess reveal that 
(1) it is continuous rather than composed of a forest of narrow emission lines, 
(2) it shows significant spectral curvature, and 
(3) it is rapidly variable.  
A simple blackbody model does not entirely fit the shape of this excess, and 
comparison with more detailed physical models is required. 
The continuous nature of the soft X-ray excess is as expected given its
rapid variability (Elvis et~al. 1991), and it parallels the recent result 
of Turner et~al. (2001) regarding the NLS1 Ton~S180. 

%--------------------------------------------------------------------------------

\section{conclusions}
We have performed the first X-ray grating spectroscopy and the first detailed 
UV absorption study of the famous Narrow-Line Seyfert~1 galaxy NGC\,4051. 
Our high-resolution X-ray and UV spectroscopic analyses reveal a number of 
significant results:
\begin{itemize}

\item{We detect a number of O, Ne, Mg and Si X-ray absorption lines belonging to 
two distinct absorption systems at blueshifted velocities of 
$-2340\pm 130$~km~s$^{-1}$ and $-600\pm 130$~km~s$^{-1}$. The former, which has 
the highest velocity of any known X-ray warm absorber in a Seyfert, is observed 
preferentially in the higher-ionization (H-like) ions, while the latter is 
stronger in lower-ionization (He-like) ions (see \S~\ref{cobs2}, \S~3.1, 
Tables~1 and 2, and Figures~\ref{megspec} and \ref{profiles}).}

\item{The STIS spectrum reveals rich intrinsic UV absorption consisting of 
multiple ($\approx 10$) 
systems, some of which are consistent in velocity with the low-velocity X-ray 
absorption system (but not the high-velocity system) and some of which are 
detected in both high-ionization (\ion{C}{4}, \ion{N}{5} and \ion{Si}{4}) and 
low-ionization (\ion{C}{2}, \ion{Si}{3} and \ion{Si}{2}) ions. Our analysis 
of archival {\it IUE} data reveals that UV absorption has been present at some 
level for more than 20~yr (see \S~\ref{stobs2}, \S~\ref{stobs3}, \S~3.1, 
Table~\ref{uvlines}, and Figures~\ref{uvspec}, \ref{uvsys}, \ref{lowsys} and 
\ref{new_uv_fig}).}

\item{The low-velocity X-ray absorption system is resolved by the MEG. We find 
it to have a FWHM of $750^{+400}_{-280}$~km~s$^{-1}$. Comparison with the UV 
absorption in the same velocity range suggests that the X-ray absorption systems 
may have complex, unresolved substructure (see \S~\ref{cobs2}, \S~3.1, and 
Figures~\ref{resolved} and \ref{rescomp}).}

\item{The {\it Chandra} spectrum contains several unresolved X-ray emission 
lines, the strongest of which 
are from \ion{O}{7} and \ion{Ne}{9}. These lines have velocities consistent 
with rest-frame emission in NGC\,4051 (see \S~\ref{cobs2}, \S~3.1, 
Table~2, and Figures~\ref{megspec} and \ref{emission}).}

\item{We estimate that the X-ray warm absorber has a column density of 
$N_{\rm H} \sim 10^{21}$~cm$^{-2}$ in the high-velocity system and 
$N_{\rm H} \sim 10^{20}$~cm$^{-2}$ in the low-velocity system 
(although we note that these values may be subject to systematic 
uncertainties). Plasma diagnostic techniques reveal that the gas is primarily 
photoionized (rather than collisionally ionized) and set upper limits of 
$\approx 10^6$~K and $\approx 4 \times 10^{10}$~cm$^{-3}$ on its 
temperature and density, respectively (see \S~3.2).}

\item{We find that the soft X-ray excess is rapidly variable, that it is continuous 
rather than made up of many narrow emission lines, and that it has significant 
spectral curvature (see \S~\ref{cobs3}, \S~3.3, Table~\ref{chivals}, and 
Figure~\ref{soft_excess}).}

\item{During the last $\approx 15$~ks of our {\it Chandra} observation, NGC\,4051 
entered a low-flux state comparable to the lowest states in which it has been 
previously observed. In the low state the spectrum became substantially harder, as 
in previously documented cases (see \S~\ref{cobs3}, \S~3.3, and 
Figures~\ref{curve} and \ref{soft_excess}).}

\end{itemize}

%--------------------------------------------------------------------------------

\acknowledgments

We thank H.~Netzer, B.~M.~Peterson, and P.~Uttley 
for helpful discussions and sharing data. 
We thank all the members of the {\it Chandra} team for their enormous
efforts.  This research is based on observations made with the NASA/ESA
{\it Hubble Space Telescope}. We gratefully acknowledge the financial
support of CXC grant GO0--1160X (MJC, WNB), STScI grant GO--08321.01--A
(MJC, WNB), the Barry Goldwater Foundation (MJC), 
the Alfred~P. Sloan Foundation (WNB), NASA LTSA grant NAG5--8107 
(SK), Hubble Fellowship grant HF-01113.01-98A (CSR), and STScI
grant GO--08321.03--A (BJW).
Hubble Fellowship grant HF-01113.01-98A was awarded by the Space Telescope 
Institute, which is operated by the Association of Universities for Research 
in Astronomy, Inc., for NASA under contract NAS 5-26555.

%--------------------------------------------------------------------------------

\end{document}